\documentclass[12pt]{article}

\usepackage{jheppub, bm, wrapfig,float,array}
\usepackage[utf8]{inputenc}
\numberwithin{equation}{section}
\renewcommand\afterTocRuleSpace{\bigskip}

\usepackage{amsfonts}
\usepackage{amsmath}
\usepackage{color}
\usepackage{graphicx}
\usepackage{float}
\usepackage[T1]{fontenc}
\usepackage[utf8]{inputenc}
\usepackage{alphabeta}
\usepackage{fancyvrb}
\usepackage[dvipsnames]{xcolor}
\usepackage{bold-extra}
\usepackage{lmodern}

\usepackage{tikz}
\usetikzlibrary{arrows}
\usetikzlibrary{shapes.geometric,calc,arrows, positioning,shapes.misc,decorations.markings}
\tikzset{
  big arrow/.style={
    decoration={markings,mark=at position 1 with {\arrow[scale=2,#1]{>}}},
    postaction={decorate},
    shorten >=0.4pt},
  big arrow/.default=black}

\newcommand{\bea}{\begin{eqnarray}}
\newcommand{\eea}{\end{eqnarray}}
\newcommand{\be}{\begin{equation}}
\newcommand{\ee}{\end{equation}}
\newcommand{\bit}{\begin{itemize}}
\newcommand{\eit}{\end{itemize}}
\newcommand{\ben}{\begin{enumerate}}
\newcommand{\een}{\end{enumerate}}

\newcommand{\half}{\frac{1}{2}}

\newcommand{\Z}{{\mathbb Z}}

\newcommand{\C}{{\mathbb C}}

\renewcommand{\P}{{\mathbb P}}

\newcommand{\cG}{\mathcal{G}}

\newcommand{\cI}{\mathcal{I}}

\newcommand{\cN}{\mathcal{N}}

\newcommand{\cS}{\mathcal{S}}

\newcommand{\F}{\mathsf{F}}
\renewcommand{\S}{\mathsf{S}}

\renewcommand{\C}{\mathsf{C}}
\newcommand{\A}{\mathsf{A}}

\renewcommand{\C}{\mathsf{C}}

\renewcommand{\L}{\mathsf{\Lambda}}

\newcommand{\fT}{\mathfrak{T}}

\newcommand{\fe}{\mathfrak{e}}
\newcommand{\ff}{\mathfrak{f}}
\newcommand{\fg}{\mathfrak{g}}

\newcommand{\su}{\mathfrak{su}}
\renewcommand{\sp}{\mathfrak{sp}}
\newcommand{\so}{\mathfrak{so}}
\renewcommand{\u}{\mathfrak{u}}

\newcommand{\ubf}[1]{\underline{\bf #1}}

\newcommand{\bF}{{\mathbb F}}

\title{Flavor Symmetry of $5d$ SCFTs, Part 2: Applications}

\author{Lakshya Bhardwaj}

\affiliation{Mathematical Institute, University of Oxford,\\Andrew Wiles Building, Woodstock Road, Oxford, OX2 6GG, UK}

\abstract{In Part 1 of this series of papers, we described a general method for determining the flavor symmetry of any $5d$ SCFT which can be constructed by integrating out BPS particles from some $6d$ SCFT compactified on a circle. In this part, we apply the method to explicitly determine the flavor symmetry of those $5d$ SCFTs which reduce, upon a mass deformation, to some $5d$ $\cN=1$ gauge theory carrying a simple gauge algebra. In these cases, the flavor symmetry of the $5d$ gauge theory is often enhanced at the conformal point. We use our method to determine this enhancement.
}

\begin{document}

\maketitle

\section{Introduction} \label{I}
In this series of papers (Part 1 \cite{Bhardwaj:2020ruf} and Part 2), we study the flavor symmetry algebras of $5d$ SCFTs\footnote{See \cite{Morrison:1996xf,Intriligator:1997pq,Diaconescu:1998cn,DelZotto:2017pti,Xie:2017pfl,Closset:2018bjz,Jefferson:2018irk,Apruzzi:2018nre,Bhardwaj:2018yhy,Bhardwaj:2018vuu,Bhardwaj:2019fzv,Apruzzi:2019vpe,Apruzzi:2019opn,Apruzzi:2019enx,Bhardwaj:2019jtr,Saxena:2019wuy,Bhardwaj:2019xeg,Apruzzi:2019syw,Bhardwaj:2020gyu,Eckhard:2020jyr,Closset:2020scj,Hubner:2020uvb,Bhardwaj:2020kim} for a study of $5d$ SCFTs by constructing them string theory compactification on singular geometries; \cite{Seiberg:1996bd,Aharony:1997ju,Aharony:1997bh,DeWolfe:1999hik,Brandhuber:1999yo,Bergman:2013aca,Zafrir:2014ywa,Zafrir:2015ftn,Hayashi:2015zka,Hayashi:2015fsa,Bergman:2015dpa,Hayashi:2018lyv,Hayashi:2018bkd,Hayashi:2019yxj} for a study of $5d$ SCFTs by constructing them through intersecting brane configurations in string theory; and see \cite{Bergman:2012rgz,DHoker:2016wak,DHoker:2017prl,DHoker:2017muf,DHoker:2017gcu,Chaney:2018gjc,Bah:2018lyv,Uhlemann:2019ypp,Uhlemann:2019ors} for their study from the point of view of holography. See also \cite{Witten:1996qb,Kim:2012gu,Zafrir:2015rga,Hayashi:2015vhy,Kim:2015jba,Ohmori:2015pia,Yonekura:2015ksa,Zafrir:2015uaa,Tachikawa:2015mha,Hayashi:2016abm,Ohmori:2016shy,Jefferson:2017ahm,Mekareeya:2017jgc,Ashok:2017bld,Bastian:2018fba,Assel:2018rcw,Bhardwaj:2019ngx,Closset:2019mdz,Hayashi:2020sly,Morrison:2020ool,Bhardwaj:2020phs,BenettiGenolini:2020doj} for other related studies. }. In Part 1 \cite{Bhardwaj:2020ruf}, we provide a general recipe for computing the flavor symmetry of any $5d$ SCFT that can be obtained (on its extended Coulomb branch) by integrating out BPS particles from the extended Coulomb branch of a known $5d$ KK theory\footnote{We define a $5d$ KK theory to be a theory obtained by compactifying a $6d$ SCFT on a circle of finite non-zero radius, possibly with twists by discrete global symmetries of the $6d$ SCFT as one traverses the circle.}. This is done by utilizing the construction of the extended Coulomb branch of a $5d$ KK theory in terms of M-theory compactified on Calabi-Yau threefolds (CY3) \cite{Bhardwaj:2018yhy,Bhardwaj:2018yhy,Bhardwaj:2019fzv,Bhardwaj:2020kim}. The flavor symmetry of this $5d$ KK theory is encoded in terms of $\P^1$ fibered non-compact surfaces coupled to the compact surfaces inside the CY3. The RG flows associated to integrating out BPS particles lead to the decoupling of some of the non-compact surfaces, leading to a new set of non-compact surfaces which encodes the flavor symmetry of the resulting $5d$ SCFT.

In this part, we apply the method discussed in Part 1 \cite{Bhardwaj:2020ruf} to explicitly determine the flavor symmetry of $5d$ SCFTs which reduce upon a mass deformation to a $5d$ $\cN=1$ gauge theory with a simple gauge algebra, and can be obtained by integrating out matter from a $5d$ KK theory. See \cite{Bhardwaj:2020gyu} for the list of all such $5d$ SCFTs which are known to exist at the time of writing of this paper.

So, consider a $5d$ SCFT $\fT$ which admits a mass deformation to a $5d$ $\cN=1$ gauge theory $\cG$. Let $\cG$ carry a semi-simple gauge algebra $\fg$ with matter content being organized as $n_i$ copies of hypermultiplets transforming in some irrep $R_i$ of $\fg$. Then, there is a classically visible flavor symmetry algebra $\ff_\cG$ that we can assign to $\cG$. If $R_i$ is a complex representation, then we obtain a factor of $\u(n_i)$ in $\ff_\cG$. If $R_i$ is a strictly real representation, then we obtain a factor of $\sp(n_i)$ in $\ff_\cG$. If $R_i$ is a pseudo-real representation, then $n_i$ is half-integral and we obtain a factor of $\so(2n_i)$ in $\ff_\cG$. Moreover, for each simple gauge algebra $\fg_a$ appearing in the semi-simple gauge algebra $\fg=\oplus_a\fg_a$, we obtain an additional $\u(1)_a$ factor in the flavor symmetry algebra whose current is provided by the instanton number for $\fg_a$. One might then wonder whether the full flavor symmetry algebra $\ff_\fT$ of $\fT$ is the same as $\ff_\cG$. It is well-known that this is not the case. In general, $\ff_\cG$ is only a subalgebra of $\ff_\fT$, but an important point is that the rank of $\ff_\cG$ equals the rank of $\ff_\fT$. This is usually stated by saying that the classical flavor symmetry $\ff_\cG$ of $\fT$ is enhanced to $\ff_\fT$ at the superconformal point, and $\ff_\fT$ is then referred to as \emph{enhanced flavor symmetry}. A classic example of enhanced flavor symmetry is provided by the Seiberg $E_n$ (where $n\le 8$) theories \cite{Seiberg:1996bd,Morrison:1996xf,Intriligator:1997pq} which admit a mass deformation to $\su(2)$ gauge theory with $n-1$ full hypers in fundamental representation. The classical flavor symmetry $\ff_\cG=\so(2n-2)\oplus\u(1)$ which is known to enhance for $n\ge2$ to $\ff_\fT=\fe_n$ where $\fe_5:=\so(10)$, $\fe_4:=\su(5)$, $\fe_3:=\su(3)\oplus\su(2)$ and $\fe_2:=\su(2)\oplus\u(1)$.

We emphasize that the method for determining the flavor symmetry of a $5d$ SCFT described in Part 1 does \emph{not} depend on the existence of a mass deformation reducing the $5d$ SCFT to a $5d$ gauge theory. That is, our method always captures the full enhanced flavor symmetry $\ff_\fT$ of the $5d$ SCFT $\fT$. In this part, we use our method to tabulate the $5d$ gauge theories with simple gauge algebra whose (associated classical) flavor symmetries are enhanced when they are UV completed into a $5d$ SCFT (where the precise meaning of the UV completion has been discussed above). See Section \ref{results} for a quick reference list of such gauge theories, where we have arranged the gauge theories according to the rank of their gauge algebra. The detailed derivation of these results has been provided in the following Section \ref{details}.

Throughout this paper, we use notation and background about geometric constructions and $5d$ KK theories that can be found in Section 5 and Appendix A of \cite{Bhardwaj:2019fzv}. We use some notation about $\P^1$ fibered surfaces that can be found in Section 4.1 of Part 1. Background and notation about geometric construction of $5d$ $\cN=1$ gauge theories can be found in Section 2 of \cite{Bhardwaj:2019ngx} and Section 3.2 of \cite{Bhardwaj:2020gyu}. Background on flops can be found in \cite{Bhardwaj:2019jtr}.

\section{Flavor symmetry of $5d$ SCFTs: Summary of results}\label{results}
In this section, we collect our results for flavor symmetry of $5d$ SCFTs that admit a mass deformation to a $5d$ $\cN=1$ gauge theory carrying a \emph{simple} gauge algebra. These flavor symmetry of a subset of these theories has been studied from other points of view in \cite{Apruzzi:2019vpe,Apruzzi:2019opn,Apruzzi:2019enx,Apruzzi:2019syw,Eckhard:2020jyr,Hubner:2020uvb,Kim:2012gu,Bergman:2013aca,Zafrir:2014ywa,Zafrir:2015ftn,Hayashi:2015fsa,Bergman:2015dpa,Hayashi:2019yxj,Hayashi:2015vhy,Yonekura:2015ksa,Zafrir:2015uaa,Tachikawa:2015mha,Jefferson:2017ahm,Zafrir:2015rga} and our results agree with the analysis of those papers.

We will denote such theories as
\be
\fg+\sum_i n_i R_i
\ee
where $\fg$ is the simple gauge algebra and $n_iR_i$ denotes that the theory contains $n_i$ hypermultiplets in irreducible representation $R_i$ of $\fg$. To account for half-hypermultiplets, we allow $n_i$ to be half-integral for pseudo-real representations. We will further abbreviate the names of various irreducible representations as follows:
\bit
\item $\F$ denotes the fundamental representations for $\su(n)$ and $\sp(n)$, the vector representation for $\so(n)$, and irreducible representations of dimensions $\mathbf{7},\mathbf{26},\mathbf{27},\mathbf{56}$ for $\fg_2,\ff_4,\fe_6,\fe_7$ respectively.
\item $\A$ denotes the adjoint representation.
\item $\L^n$ denote the irreducible $n$-index antisymmetric representations for $\su(n)$ and $\sp(n)$.
\item $\S^2$ denotes the 2-index symmetric representation for $\su(n)$.
\item $\S$ denotes irreducible spinor representation for $\so(n)$.
\item $\C$ denotes irreducible co-spinor representation for $\so(2n)$.
\eit
Furthermore, for $\fg=\su(n)$ we have to specify a Chern-Simons level\footnote{In this paper, we adopt the convention that the Chern-Simons level is captured by a tree-level contribution (related to the Cubic casimir) to the prepotential of the $5d$ $\su(n)$ gauge theory.} $k$, which we include as a subscript of $\su(n)$, and describe such a theory as $\su(n)_k+\sum_i n_i R_i$. For $\sp(n)$ we sometimes have to specify a theta angle $\theta$ which can take values $0,\pi$ only, and we describe such a theory as $\sp(n)_\theta+\sum_i n_i R_i$.

The list of $5d$ gauge theories with simple gauge algebra that are known to UV complete to $5d$ SCFTs has been compiled in \cite{Bhardwaj:2020gyu}, to which we refer the reader. The only gauge theories in their list which cannot be obtained from $5d$ KK theories by integrating out BPS particles are as follows \cite{Bhardwaj:2020gyu,Bhardwaj:2019xeg}:
\bit
\item $\ff_4+n\F$ for $1\le n\le 3$.
\item $\fe_6+n\F$ for $1\le n\le 4$.
\item $\fe_7+\frac n2\F$ for $1\le n\le 6$.
\eit
In this section, we provide the flavor symmetry of all $5d$ SCFTs appearing in \cite{Bhardwaj:2020gyu} except for the three kinds of theories listed above.

We will use either $\fT$ or $\fT_n$ to denote the theories and $\ff(\fT)$ or $\ff(\fT_n)$ to denote their flavor symmetries. Some $5d$ SCFTs can reduce to multiple $5d$ gauge theories (with a simple gauge algebra) if one deforms them by different mass parameters. In this case, one says that the different $5d$ gauge theory descriptions are related by \emph{$5d$ dualities}. Below, we account for such dualities by placing an `=' sign between the different $5d$ gauge theory descriptions. For example, the $5d$ SCFTs appearing in (\ref{t1}) have two gauge theory descriptions; one of them being $\su(m+2)_\frac n2+(2m+8-n)\F$, and the other being $\sp(m+1)+(2m+8-n)\F$.

Below, we will only mention theories for which there is a \emph{non-trivial enhancement} of flavor symmetry at the conformal point. The flavor symmetry for theories not being mentioned in this section, but appearing in \cite{Bhardwaj:2020gyu}, is simply the classical flavor symmetry associated to the gauge theory. As an example, for $n=2m+7$ and $n=2m+8$ in (\ref{t1}) there is no enhancement of classical flavor symmetry, and hence those cases are omitted. On the other hand, some of the gauge theories have an enhancement that is visible from the viewpoint of a dual gauge theory. Such cases are \emph{not} omitted below. An example of such a case is (\ref{t1}) for $3\le n\le 2m+6$.

\subsection{General Rank}
\be\label{t1}

\ee

\subsection{Rank 1}
\be\label{t20}
%
\ee

\subsection{Rank 2}
\be\label{t22}
%
\ee

\subsection{Rank 3}
\be\label{t25}
%
\ee

\subsection{Rank 4}
\be\label{t46}
%
\ee

\subsection{Rank 5}
\be\label{t60}
%
\ee

\subsection{Rank 6}
\be\label{t85}
%
\ee

\subsection{Rank 7}
\be\label{t99}
%
\ee

\section{Detailed analysis}\label{details}
\subsection{General Rank}
\subsubsection*{\ubf{Derivation of (\ref{t1})}:}
Let us start with the derivation of (\ref{t1}). The theories $\sp(m+1)+(2m+8-n)\F$ can be obtained from
\be\label{e1}
\sp(m+1)+(2m+8)\F
\ee
by integrating out fundamental hypers. It is known that the $5d$ $\cN=1$ gauge theory (\ref{e1}) is a $5d$ KK theory and can be obtained by an untwisted circle compactification of the $6d$ SCFT whose tensor branch description is provided by the $6d$ $\cN=(1,0)$ gauge theory $\sp(m)+(2m+8)\F$. We denote this fact by an equation of the following form
\be\label{e3}

\ee
where the notation for $5d$ KK theories is borrowed from \cite{Bhardwaj:2019fzv}. According to \cite{Bhardwaj:2019ngx,Bhardwaj:2020gyu}, the above equality can be seen geometrically as follows. Consider the resolved CY3 geometry described by
\be\label{g1}
%
\ee
which describes the $5d$ KK theory
\be
%
\ee
Now applying the isomorphism $\cS$ (which exchanges $e$ and $f$ in a surface $\bF_0^b$) on the left-most surface of (\ref{g1}) leads to the geometry
\be
%
\ee
which describes the $5d$ gauge theory $\sp(m+1)+(2m+8)\F$. This isomorphism establishes (\ref{e3}).

Since the $6d$ SCFT
\be
%
\ee
has a $\so(4m+16)$ flavor symmetry, we expect to be able to couple the geometry (\ref{g1}) to a collection of non-compact $\P^1$ fibered surfaces $\mathbf{N}_i$ such that their associated intersection matrix\footnote{The intersection matrix is defined as $-f_i\cdot \mathbf{N}_j$ where $f_i$ is the $\P^1$ fiber of $\mathbf{N}_i$.} gives rise to the Cartan matrix for the affine Lie algebra $\so(4m+16)^{(1)}$. According to the gluing rules, this coupling takes the following form
\be\label{g3}
%
\ee
where $\mathbf{N}_i$ denote\footnote{Unlike the case for compact surfaces, the subscript $i$ for non-compact surfaces $\mathbf{N}_i$ should not be interpreted as the ``degree'' of the surface. It is simply a labeling of the non-compact surfaces.} the non-compact surfaces corresponding to $\so(4m+16)^{(1)}$. The $e$ curves living in $\mathbf{N}_i$ are non-compact sections whose crucial property is that 
\be\label{e4}
e\cdot f=1
\ee
We emphasize that any section in $\mathbf{N}_i$ satisfying (\ref{e4}) is being denoted by $e$ in our notation. Correspondingly different appearances of $e$ for a single non-compact surfaces should be regarded as two different sections which may not even be in the same homology class inside the surface. For example, there are three such sections of $\mathbf{N}_2$ appearing in (\ref{g3}), namely the curves gluing $\mathbf{N}_2$ to $\mathbf{N}_0$, $\mathbf{N}_1$ and $\mathbf{N}_3$. Despite all these three sections being denoted by $e$, these three sections should be understood as three different sections without any apriori relationship between their homology classes inside $\mathbf{N}_2$.

Performing $\cS$ on $\bF_0^{2m+8}$ converts (\ref{g3}) into
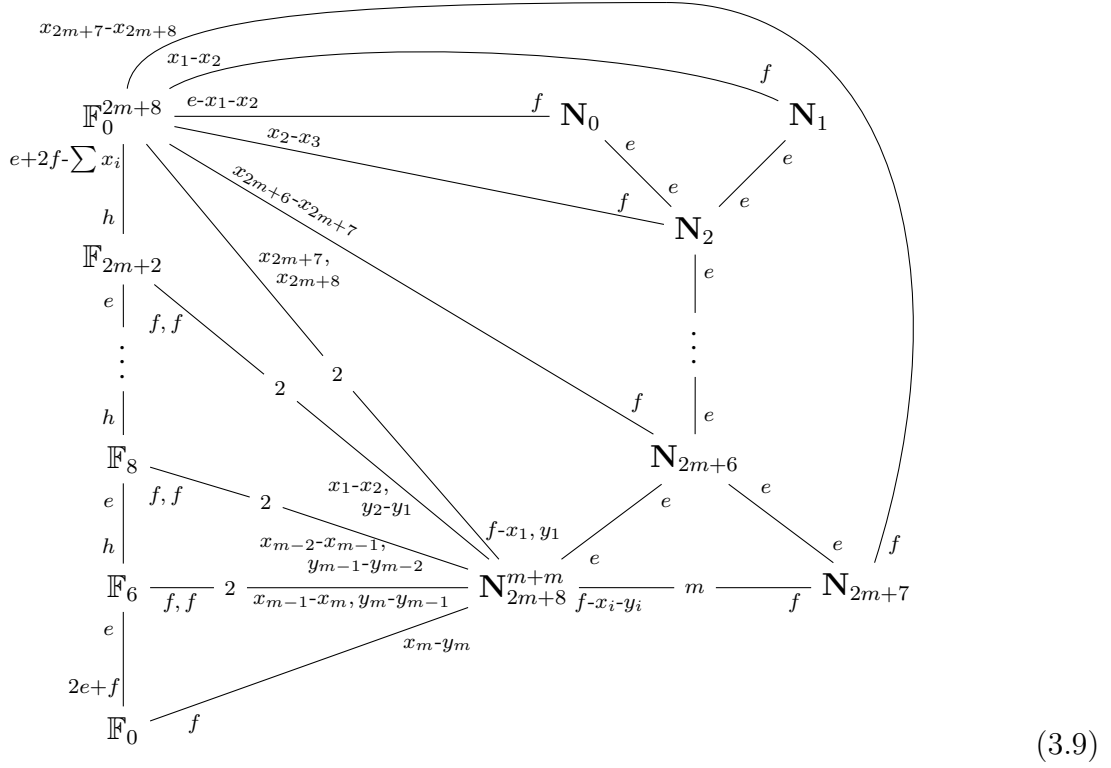
\be\label{g2}

\ee
Now to integrate out an $\F$ of $\sp(m+1)$, we have to first flop the curve $f-x_1$ living in $\bF_0^{2m+8}$ of (\ref{g2}) to obtain the following geometry
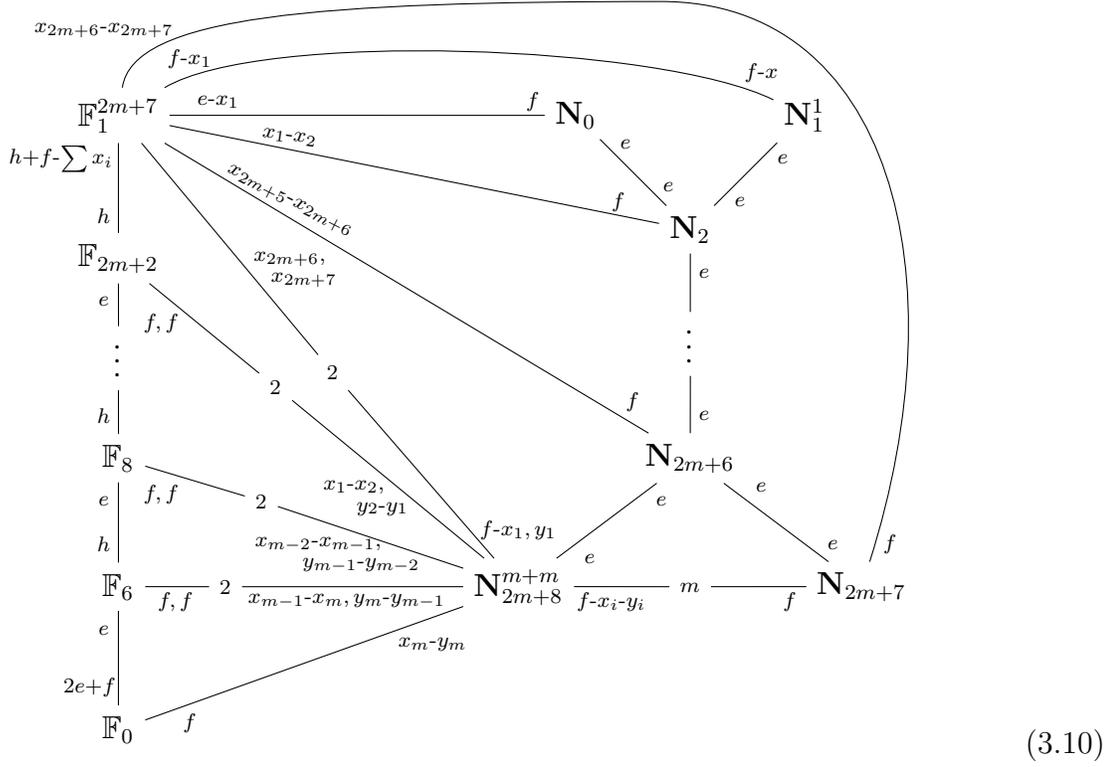
\be
%
\ee
where we have relabeled the blowups living in the resulting surface $\bF_1^{2m+7}$. The flopped curve can be identified with the blowup $x$ living in $\mathbf{N}_1$. To complete the process of integrating out of the flavor, we have to expand this blowup $x$ to infinite volume while keeping all the curves living in the compact surfaces at finite volume. In particular, we need to keep the curve $f-x_1$ living in $\bF_1^{2m+7}$, which, since it is identified with the curve $f-x$ living in $\mathbf{N}_1$, implies that the curve $f$ living in $\mathbf{N}_1$ must go to infinite volume as well. Thus, the $\P^1$ fibration of the non-compact surface $\mathbf{N}_1$ is destroyed once we integrate out the flavor. After this process, we obtain the following geometry comprised of compact surfaces and $\P^1$ fibered non-compact surfaces
\be\label{g4}

\ee
which implies that $\sp(m+1)+(2m+7)\F$ carries an $\so(2m+16)$ flavor symmetry, as can be seen by computing the intersection matrix of the remaining $\P^1$ fibered non-compact surfaces in the above geometry.

Now, removing another flavor corresponds to flopping $f-x_1$ living inside $\bF_1^{2m+7}$ of (\ref{g4}). This leads to the following geometry
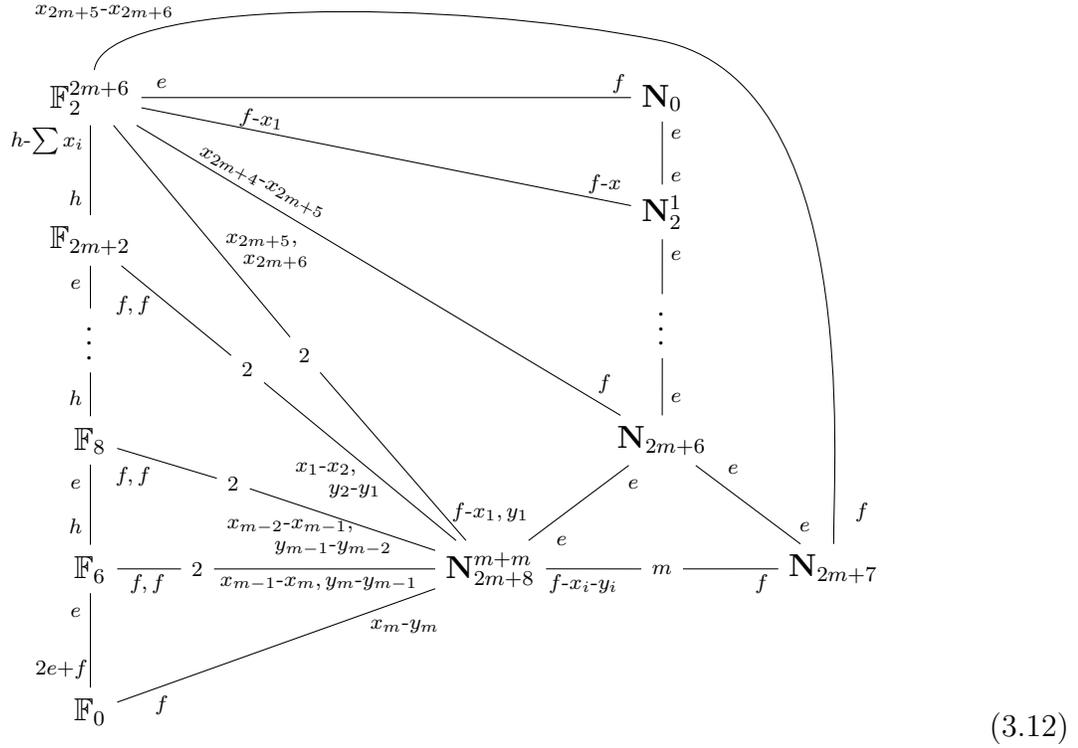
\be
%
\ee
where we have again relabeled the blowups on the resulting surface $\bF_2^{2m+6}$. By similar argument as above, sending the volume of the blowup $x$ living in $\mathbf{N}_2$ to infinity decouples the surface $\mathbf{N}_2$, and we are left with the geometry
\be\label{g5}
%
\ee
implying that the flavor symmetry for $\sp(m+1)+(2m+6)\F$ is $\so(2m+12)\oplus\su(2)$.

At the next step, an interesting phenomenon occurs. Integrating out another flavor corresponds to flopping $f-x_1$ living in $\bF_2^{2m+6}$ of (\ref{g5}) and leads to
\be
%
\ee
Thus, integrating out this flavor decouples \emph{two} non-compact surfaces namely $\mathbf{N}_0$ and $\mathbf{N}_3$, thus reducing the rank of the non-abelian part of the flavor symmetry by two. However, since we have only integrated out a single flavor, the rank of the full flavor symmetry algebra should only reduce by one. This implies that a $\u(1)$ factor should arise in the full flavor symmetry algebra of the resulting theory. That is, the flavor symmetry for $\sp(m+1)+(2m+5)\F$ should be $\so(2m+10)\oplus\u(1)$. In this paper, we are not going to track $\u(1)$ factors in the geometry, but instead track them by matching the rank of the non-abelian part of the flavor symmetry (as deduced from geometry) with the rank of the full flavor symmetry, in order to obtain the number of missing $\u(1)$ factors.

Continuing in this fashion we observe that the geometry for $\sp(m+1)+\F$ contains no non-compact $\P^1$ fibered surfaces. Consequently, the geometry for pure $\sp(m+1)_\theta$ won't contain any non-compact $\P^1$ fibered surfaces, irrespective of the value of $\theta$. Thus, the flavor symmetry for $\sp(m+1)_\theta$ with $m\ge1$ is $\u(1)$ for $\theta=0,\pi$.

\subsubsection*{\ubf{Derivation of (\ref{t2})}:}
To produce theories listed in (\ref{t2}), we start with
\be\label{e2}

\ee
which is implemented by doing an $\cS$ transformation on both $\bF_0^{2m+8}$ and $\bF_0$ in (\ref{g3}), which gives
\be
%
\ee
Now we flop the blowup $x_{2m+8}$ living in the top-most compact surface $\bF_0^{2m+8}$ to the bottom-most compact surface $\bF_0$ to rewrite the above geometry as
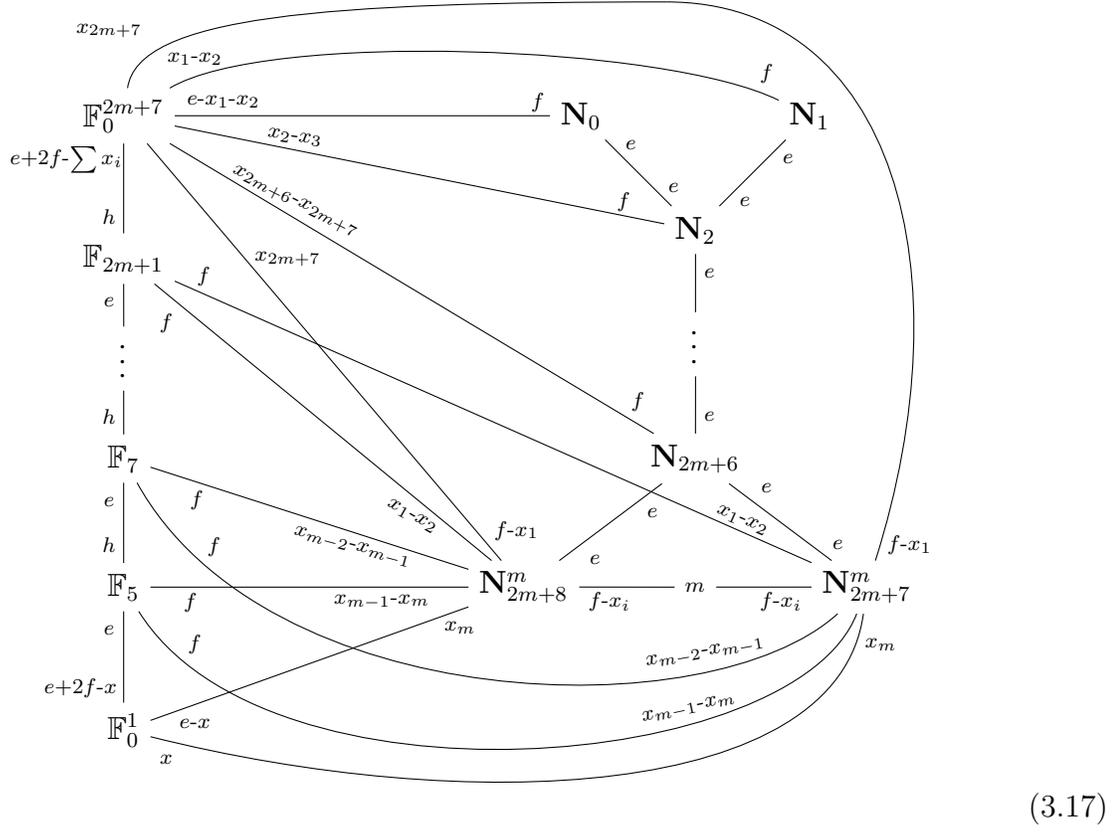
\be
%
\ee
The theory $\su(m+2)_0+(2m+6)\F$ is produced by flopping $f-x$ living in $\bF_0^1$ and $f-x_1$ living in $\bF_0^{2m+7}$ out of the geometry. This leads to the geometry
\be\label{g6}
%
\ee
The reader can verify in the same way as above that demanding all curves inside compact surfaces to have finite volume implies that $f$ of $\mathbf{N}_{2m+7}$ goes to infinite size. According to the above geometry, we find that the flavor symmetry for $\su(m+2)_0+(2m+6)\F$ is $\su(2m+8)$. Subsequent theories in (\ref{t2}) are produced by flopping and integrating out the curves $f-x_i$ living in the top-most compact surface as discussed above for the case of (\ref{t1}).

\subsubsection*{\ubf{Derivation of (\ref{t3})}:}
Let us flop $x_{2m+6}$ from the top-most compact surface to the bottom-most compact surface in (\ref{g6}). This leads to the geometry
\be

\ee
The theory $\su(m+2)_0+(2m+4)\F$ is produced by integrating out  $f-x$ in $\bF_1^1$ and $f-x_1$ in $\bF_1^{2m+5}$. Other theories in (\ref{t3}) are produced by successively integrating out $f-x_i$ from the top-most compact surface. The reader can easily check that integrating out these curves leads precisely to the results mentioned in (\ref{t3}). The reader can also check that the theories
\be
\su(m+2)_\frac{n-1}2+(2m+5-2p-n)\F
\ee
for $m,n,p\ge1$ that can also be produced by integrating out matter from (\ref{e2}) have no enhancement of flavor symmetry.

\subsubsection*{\ubf{Derivation of (\ref{t4}) and (\ref{t5})}:}
We can produce these theories by integrating out fundamental matter from the KK theory
\be\label{e6}

\ee
The flavor symmetry for the $6d$ SCFT
\be
%
\ee
is $\su(m+8)\oplus\u(1)$ as long as $m\ge5$. Let us reproduce the geometry for
\be
%
\ee
which manifests the coupling to a collection of non-compact surfaces with intersection matrix $\su(m+8)^{(1)}$
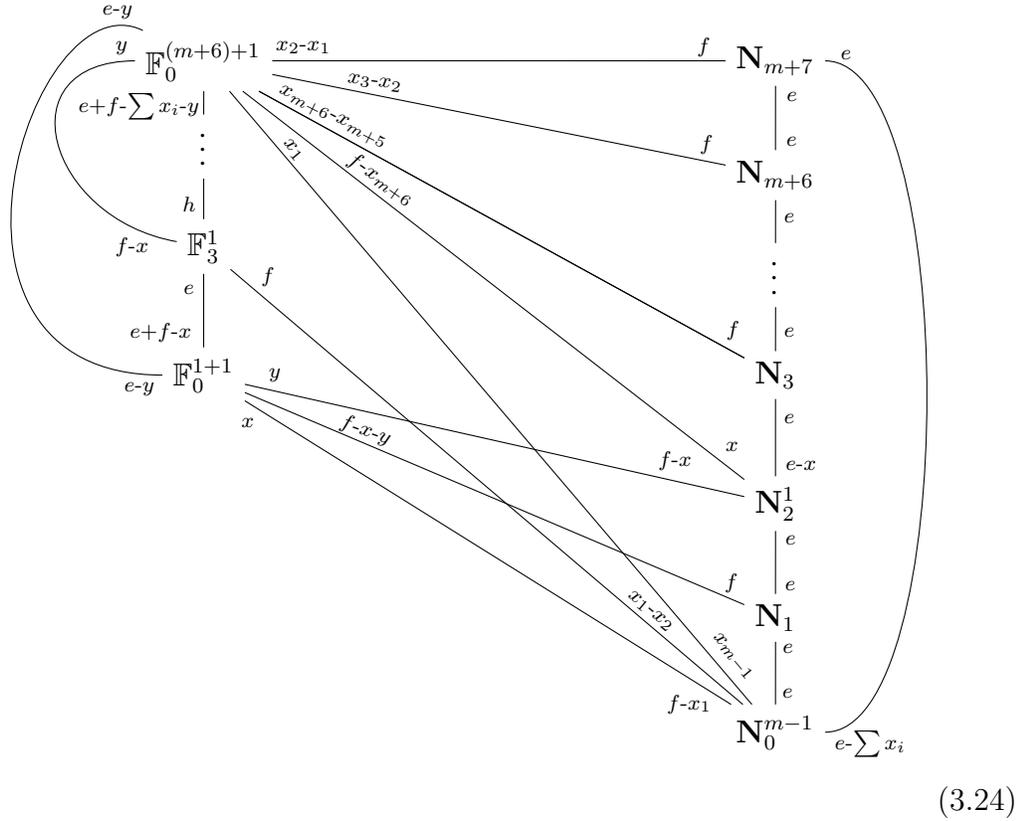
\be
%
\ee
The geometry for $5d$ theory
\be
\su(m+1)_0+\L^2+(m+7)\F
\ee
is produced by applying $\cS$ on the top-most and bottom-most compact surfaces
\be\label{g7}
%
\ee
The theories in (\ref{t4}) are produced by integrating out curves $f-x_i$ from the top-most compact surface of the above geometry. The first step corresponds to integrating out $f-x_{m+6}$ and we can see that it destroys the $\P^1$ fibration of the surface $\mathbf{N}_3$ thus leading to an $\su(m+8)$ non-abelian part of the flavor symmetry. Combining it with the extra $\u(1)$ flavor symmetry descending from the $6d$ SCFT we find that the flavor symmetry for $\su(m+1)_\half+\L^2+(m+6)\F$ is $\u(m+8)$, as claimed in (\ref{t4}). The reader can similarly check the remaining claims in (\ref{t4}).

The theories in (\ref{t5}) can be produced by first integrating out $f-x$ from the bottom-most compact surface followed by integrating out the curves $f-x_i$ from the top-most compact surface in (\ref{g7}).

Finally, note that we have only derived (\ref{t5}) for $m\ge5$. For $m=4$, we will derive it in Section \ref{R4d}.

\subsubsection*{\ubf{Derivation of (\ref{t6}---\ref{te1})}:}
This class of theories can be produced by integrating out matter from the KK theory
\be
\begin{tikzpicture} [scale=1.9]
\node at (-5.7,0.9) {$\sp(m+1)+\L^2+8\F$};
\node at (-4.5,0.9) {$=$};
\node (v1) at (-1.5,0.8) {2};
\node at (-1.5,1.1) {$\su(1)^{(1)}$};
\node at (-8.3,0.9) {$\su(m+2)_{\frac m2+1}+\L^2+8\F$};
\node at (-6.9,0.9) {$=$};
\node (v3) at (-2.9,0.8) {2};
\node at (-2.9,1.1) {$\su(1)^{(1)}$};
\node (v2) at (-2.2,0.8) {$\cdots$};
\draw  (v2) edge (v3);
\draw  (v2) edge (v1);
\node (v4) at (-3.8,0.8) {1};
\node at (-3.8,1.1) {$\sp(0)^{(1)}$};
\draw  (v4) edge (v3);
\begin{scope}[shift={(0,0.45)}]
\node at (-2.2,-0.15) {$m$};
\draw (-3.1,0.15) .. controls (-3.1,0.1) and (-3.1,0.05) .. (-3,0.05);
\draw (-3,0.05) -- (-2.3,0.05);
\draw (-2.2,0) .. controls (-2.2,0.05) and (-2.25,0.05) .. (-2.3,0.05);
\draw (-2.2,0) .. controls (-2.2,0.05) and (-2.15,0.05) .. (-2.1,0.05);
\draw (-2.1,0.05) -- (-1.4,0.05);
\draw (-1.3,0.15) .. controls (-1.3,0.1) and (-1.3,0.05) .. (-1.4,0.05);
\end{scope}
\end{tikzpicture}
\ee
The corresponding $6d$ SCFT has an $\fe_8\oplus\su(2)$ flavor symmetry. The $\fe_8$ factor arises from the $\sp(0)$ node and the $\su(2)$ factor is a delocalized flavor symmetry associated to the $\su(1)$ nodes. Correspondingly we expect that the compact part of the geometry for the above KK theory can be coupled to non-compact $\P^1$ fibered surfaces whose intersection matrix comprises the Cartan matrix for $\fe_8^{(1)}\oplus\su(2)^{(1)}$. We will denote the non-compact surfaces comprising $\fe_8^{(1)}$ as $\mathbf{N}_i$ and the non-compact surfaces comprising $\su(2)^{(1)}$ as $\mathbf{M}_i$. The geometry can be written as
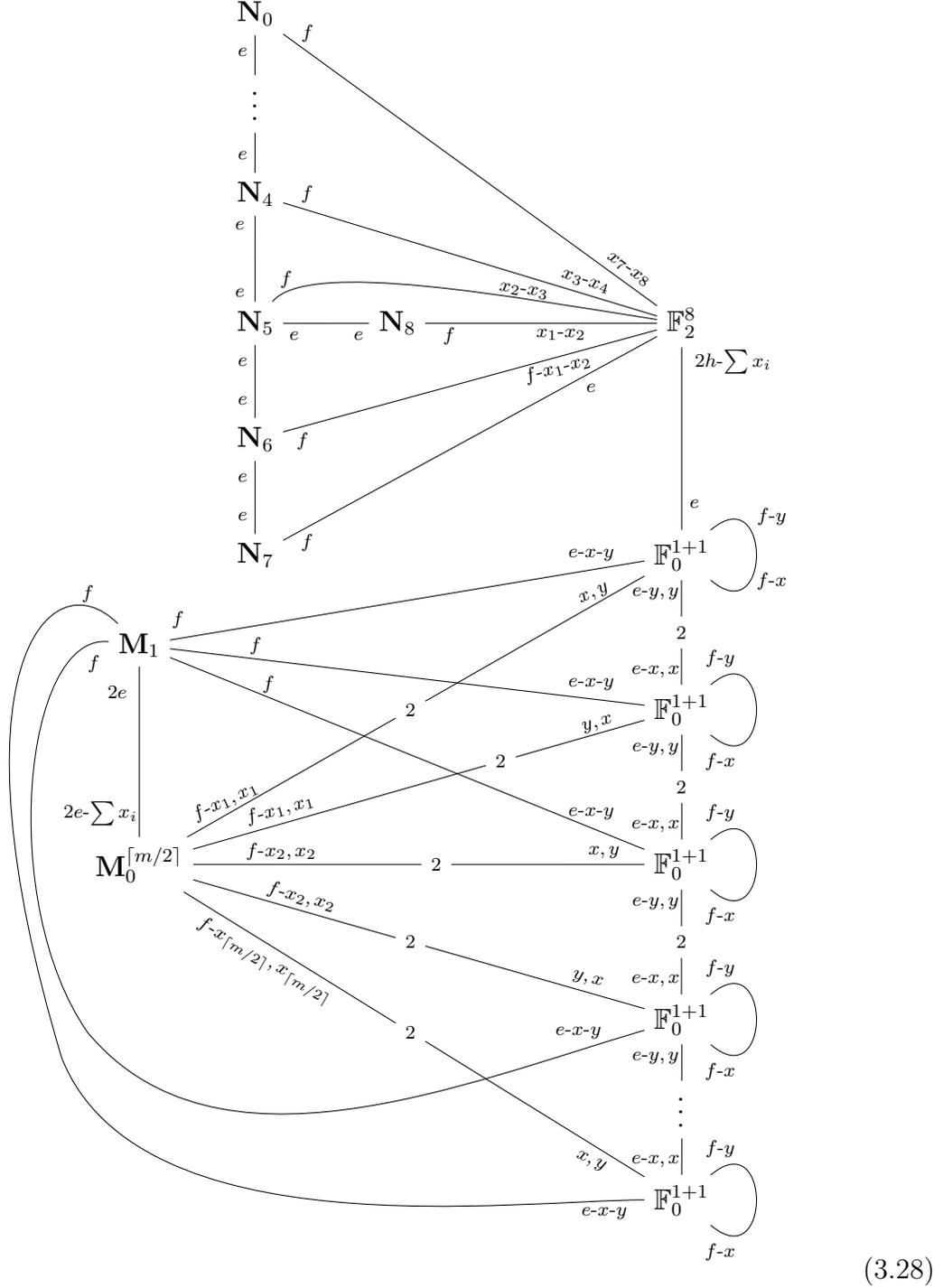
\be

\ee
which manifests the $\sp(m+1)+\L^2+8\F$ $5d$ gauge theory description of the KK theory. The theories in (\ref{t6}) can be produced by successively integrating out $x_i$ living in the top-most compact surface. It is easy to read how these flops affect the non-compact surfaces. At the first step, integrating out $x_8$ integrates out $\mathbf{N}_0$ and $\mathbf{M}_1$, thus leading to an $\fe_8\oplus\su(2)$ flavor symmetry. Subsequent flops only affect the surfaces $\mathbf{N}_i$ and so an $\su(2)$ factor is present in the flavor symmetry for all $5d$ SCFTs in this class.\\
The geometry for
\be
\su(m+2)_{\frac{m+9}2}+\L^2+\F\:=\:\sp(m+1)+\L^2+\F
\ee
can be written as
\be

\ee
Now, integrating out $x$ living in the top-most compact surface leads to the theory $\sp(m+1)_0+\L^2$, while integrating out $f-x$ living in the top-msot compact surface leads to the theory $\sp(m+1)_{\pi}+\L^2$. The former RG flow preserves both $\mathbf{N}_7$ and $\mathbf{M}_0$ while the latter RG flow only preserves $\mathbf{M}_0$, thus implying that the flavor symmetry is $\su(2)^2$ when $\theta=0$ but only $\u(2)$ when $\theta=\pi$. Combining this with the duality
\be
\su(m+2)_{\frac{m}2+5}+\L^2\:=\:\sp(m+1)_{m\pi}+\L^2
\ee
we derive the results (\ref{te0}---\ref{te1}).

\subsubsection*{\ubf{Derivation of (\ref{t7}---\ref{t11})}:}
These theories can be produced by using the KK theory
\be

\ee
The corresponding $6d$ SCFT has an $\fe_7\oplus\su(2)^3$ flavor symmetry. The $\fe_7$ arises from the $\sp(0)$ node, one $\su(2)$ arises from the two fundamental hypers situated at the left end of the chain of $\su(2)$ nodes, one $\su(2)$ arises from the two fundamental hypers situated at the right end of the chain of $\su(2)$ nodes, and one $\su(2)$ is a delocalized symmetry rotating all the bifundamentals between the $\su(2)$ nodes. For $m$ even, we write the geometry for the KK theory as
\be
%
\ee
where we have have labeled the compact surfaces as $\mathbf{i}_n^b$ which denotes $\bF_n^b$ and $\mathbf{i}$ is simply a label allowing us to refer to this surface as $\mathbf{S}_i$, which we shall do in what follows. We have also displayed all the $\P^1$ fibered non-compact surfaces. However, we have omitted all the ``mutual'' edges, that is edges between compact and non-compact surfaces, and edges between non-compact surfaces comprising different simple factors of the flavor symmetry algebra (or its affinized version). The data of these omitted edges is displayed in the following gluing rules:
\bit
\item $h-x_1-x_2-x_3-x_4$ in $\mathbf{S}_{m+1}$ is glued to $f$ in $\mathbf{N}_7$.
\item $x_i-x_{i+1}$ in $\mathbf{S}_{m+1}$ is glued to $f$ in $\mathbf{N}_{i-1}$ for $i=1,\cdots,7$.
\item $y_1,y_2$ in $\mathbf{S}_{m+2}$ are glued to $x_1,x_2$ in $\mathbf{P}_0$.
\item $e-y_1-y_2$ in $\mathbf{S}_{m+2}$ is glued to $f$ in $\mathbf{P}_1$.
\item $e$ in $\mathbf{S}_{m}$ is glued to $f-x_1-x_2$ in $\mathbf{P}_0$.
\item $x_1,x_2$ in $\mathbf{S}_{1}$ are glued to $x_1,x_2$ in $\mathbf{Q}_0$.
\item $e-x_1-x_2$ in $\mathbf{S}_{1}$ is glued to $f$ in $\mathbf{Q}_1$.
\item $e$ in $\mathbf{S}_{2m+1}$ is glued to $f-x_1-x_2$ in $\mathbf{Q}_0$.
\item $x_1-x_2,y_2-y_1$ in $\mathbf{S}_{m+2i}$ are glued to $f,f$ in $\mathbf{M}_1$ for $i=1,\cdots,\frac m2$.
\item $x_2-x_1,y_1-y_2$ in $\mathbf{S}_{m+1-2i}$ are glued to $f,f$ in $\mathbf{M}_1$ for $i=1,\cdots,\frac m2$.
\item $e-x_1,x_2,e-y_2,y_1$ in $\mathbf{S}_{m+2i}$ are glued to $f-x_{2i},y_{2i},x_{2i-1},y_{2i-1}$ in $\mathbf{M}_0$ for $i=1,\cdots,\frac m2$.
\item $e-x_2,x_1,e-y_1,y_2$ in $\mathbf{S}_{m+1-2i}$ are glued to $f-x_{2i+1},y_{2i+1},x_{2i},y_{2i}$ in $\mathbf{M}_0$ for $i=1,\cdots,\frac m2$.
\item $e,e$ in $\mathbf{S}_{m+2-2i}$ are glued to $x_{2i}-y_{2i},f-x_{2i-1}-y_{2i-1}$ in $\mathbf{M}_0$ for $i=1,\cdots,\frac m2$.
\item $e,e$ in $\mathbf{S}_{m+1+2i}$ are glued to $x_{2i+1}-y_{2i+1},f-x_{2i}-y_{2i}$ in $\mathbf{M}_0$ for $i=1,\cdots,\frac m2$.
\item $x_2-x_1$ in $\mathbf{P}_{0}$ is glued to $f$ in $\mathbf{M}_1$.
\item $f-x_2,x_1$ in $\mathbf{P}_{0}$ is glued to $f-x_1,y_1$ in $\mathbf{M}_0$.
\item $f$ in $\mathbf{P}_{1}$ is glued to $x_1-y_1$ in $\mathbf{M}_0$.
\item $x_2-x_1$ in $\mathbf{Q}_{0}$ is glued to $f$ in $\mathbf{M}_1$.
\item $f-x_2,x_1$ in $\mathbf{Q}_{0}$ is glued to $x_{m+1},y_{m+1}$ in $\mathbf{M}_0$.
\item $f$ in $\mathbf{Q}_{1}$ is glued to $f-x_{m+1}-y_{m+1}$ in $\mathbf{M}_0$.
\eit
For $m$ odd, we write the geometry for the KK theory as
\be

\ee
along with the following gluing rules
\bit
\item $h-x_1-x_2-x_3-x_4$ in $\mathbf{S}_{m+1}$ is glued to $f$ in $\mathbf{N}_7$.
\item $x_i-x_{i+1}$ in $\mathbf{S}_{m+1}$ is glued to $f$ in $\mathbf{N}_{i-1}$ for $i=1,\cdots,7$.
\item $y_1,y_2$ in $\mathbf{S}_{m+2}$ are glued to $x_1,x_2$ in $\mathbf{P}_0$.
\item $e-y_1-y_2$ in $\mathbf{S}_{m+2}$ is glued to $f$ in $\mathbf{P}_1$.
\item $e$ in $\mathbf{S}_{m}$ is glued to $f-x_1-x_2$ in $\mathbf{P}_0$.
\item $x_1,x_2$ in $\mathbf{S}_{2m+1}$ are glued to $x_1,x_2$ in $\mathbf{Q}_0$.
\item $e-x_1-x_2$ in $\mathbf{S}_{2m+1}$ is glued to $f$ in $\mathbf{Q}_1$.
\item $e$ in $\mathbf{S}_{1}$ is glued to $f-x_1-x_2$ in $\mathbf{Q}_0$.
\item $x_1-x_2,y_2-y_1$ in $\mathbf{S}_{m+2i}$ are glued to $f,f$ in $\mathbf{M}_1$ for $i=1,\cdots,\frac{m+1}2$.
\item $x_2-x_1,y_1-y_2$ in $\mathbf{S}_{m+1-2i}$ are glued to $f,f$ in $\mathbf{M}_1$ for $i=1,\cdots,\frac{m-1}2$.
\item $e-x_1,x_2,e-y_2,y_1$ in $\mathbf{S}_{m+2i}$ are glued to $f-x_{2i},y_{2i},x_{2i-1},y_{2i-1}$ in $\mathbf{M}_0$ for $i=1,\cdots,\frac{m+1}2$.
\item $e-x_2,x_1,e-y_1,y_2$ in $\mathbf{S}_{m+1-2i}$ are glued to $f-x_{2i+1},y_{2i+1},x_{2i},y_{2i}$ in $\mathbf{M}_0$ for $i=1,\cdots,\frac{m-1}2$.
\item $e,e$ in $\mathbf{S}_{m+2-2i}$ are glued to $x_{2i}-y_{2i},f-x_{2i-1}-y_{2i-1}$ in $\mathbf{M}_0$ for $i=1,\cdots,\frac{m+1}2$.
\item $e,e$ in $\mathbf{S}_{m+1+2i}$ are glued to $x_{2i+1}-y_{2i+1},f-x_{2i}-y_{2i}$ in $\mathbf{M}_0$ for $i=1,\cdots,\frac{m-1}2$.
\item $x_2-x_1$ in $\mathbf{P}_{0}$ is glued to $f$ in $\mathbf{M}_1$.
\item $f-x_2,x_1$ in $\mathbf{P}_{0}$ is glued to $f-x_1,y_1$ in $\mathbf{M}_0$.
\item $f$ in $\mathbf{P}_{1}$ is glued to $x_1-y_1$ in $\mathbf{M}_0$.
\item $x_1-x_2$ in $\mathbf{Q}_{0}$ is glued to $f$ in $\mathbf{M}_1$.
\item $f-x_1,x_2$ in $\mathbf{Q}_{0}$ is glued to $x_{m+1},y_{m+1}$ in $\mathbf{M}_0$.
\item $f$ in $\mathbf{Q}_{1}$ is glued to $f-x_{m+1}-y_{m+1}$ in $\mathbf{M}_0$.
\eit
The theories in (\ref{t7}) are produced by successively integrating out $x_i$ living in $\mathbf{S}_{m+1}$. This integrates out $\mathbf{P}_0$, $\mathbf{Q}_1$ and $\mathbf{M}_0$ for $m$ even, and $\mathbf{P}_0$, $\mathbf{Q}_0$ and $\mathbf{M}_0$ for $m$ odd. The affect on surfaces $\mathbf{N}_i$ is same in both cases. Thus the flavor symmetry takes the form $\ff\oplus\su(2)^3$ (where the subfactor $\ff$ originates from the surfaces $\mathbf{N}_i$) irrespective of whether $m$ is even or odd.

To produce theories in (\ref{t8}), we first integrate out $f-x_1$ living in $\mathbf{S}_{m+1}$, which integrates out $\mathbf{N}_1$, $\mathbf{P}_1$, $\mathbf{Q}_0$, $\mathbf{M}_0$ for $m$ even, and $\mathbf{N}_1$, $\mathbf{P}_1$, $\mathbf{Q}_1$, $\mathbf{M}_0$ for $m$ odd. Then, we successively integrate out other $x_i$ living in $\mathbf{S}_{m+1}$. The combined effect is that only $\mathbf{M}_1$ survives out of the surfaces $\mathbf{M}_i$, $\mathbf{P}_i$ and $\mathbf{Q}_i$, irrespective of whether $m$ is even or odd. The effect on $\mathbf{N}_i$ is same for both cases. Thus non-abelian part of the global symmetry takes the form $\ff\oplus\su(2)$ for all these theories.

To produce theories in (\ref{t9}), we first integrate out $f-x_1$, $f-x_2$, $x_8$ (in that order) before successively integrating out other $x_i$ living in $\mathbf{S}_{m+1}$. To produce theories in (\ref{t10}), we first integrate out $f-x_1$, $f-x_2$, $f-x_3$, $x_8$, $x_7$ (in that order) before successively integrating out other $x_i$ living in $\mathbf{S}_{m+1}$. To produce (\ref{t11}), we integrate out $f-x_1$, $f-x_2$, $f-x_3$, $f-x_4$, $x_8$, $x_7$, $x_6$, $x_5$ (in that order). In all these cases only $\mathbf{M}_1$ survives out of the surfaces $\mathbf{M}_i$, $\mathbf{P}_i$ and $\mathbf{Q}_i$. Thus the non-abelian part of the flavor symmetry takes the $\ff\oplus\su(2)$ where $\ff$ is read from the surviving $\mathbf{N}_i$.

\subsubsection*{\ubf{Derivation of (\ref{t12}---\ref{t14})}:}
These theories can be produced by using the KK theory
\be

\ee
where the corresponding $6d$ SCFT has an $\so(16)\oplus\su(2)^2$ flavor symmetry. The $\so(16)$ factor arises from eight fundamental hypers charged only under $\sp(1)$, one $\su(2)$ factor arises from the two fundamental hypers charged under the right-most $\su(2)$ node only, and the other $\su(2)$ factor corresponds to a delocalized symmetry rotating all the bifundamentals. For $m$ odd, the geometry can be written as
\be
%
\ee
along with the following gluing rules:
\bit
\item $e-x_1-x_2$ in $\mathbf{S}_{m+1}$ is glued to $f$ in $\mathbf{N}_0$.
\item $x_i-x_{i+1}$ in $\mathbf{S}_{m+1}$ is glued to $f$ in $\mathbf{N}_{i}$ for $i=1,\cdots,7$.
\item $x_7,x_8$ in $\mathbf{S}_{m+1}$ are glued to $f-x,y$ in $\mathbf{N}_8$.
\item $e$ in $\mathbf{S}_{m+2}$ is glued to $x-y$ in $\mathbf{N}_8$.
\item $x_1,x_2$ in $\mathbf{S}_{1}$ are glued to $x_1,x_2$ in $\mathbf{P}_0$.
\item $e-x_1-x_2$ in $\mathbf{S}_{1}$ is glued to $f$ in $\mathbf{P}_1$.
\item $e$ in $\mathbf{S}_{2m+2}$ is glued to $f-x_1-x_2$ in $\mathbf{P}_0$.
\item $x_1-x_2$ in $\mathbf{S}_{m+2}$ is glued to $f$ in $\mathbf{M}_1$.
\item $e-x_1,x_2$ in $\mathbf{S}_{m+2}$ are glued to $f-x_{1},y_{1}$ in $\mathbf{M}_0$.
\item $e$ in $\mathbf{S}_{m+1}$ is glued to $x_{1}-y_{1}$ in $\mathbf{M}_0$.
\item $x_2-x_1,y_1-y_2$ in $\mathbf{S}_{m+2-2i}$ are glued to $f,f$ in $\mathbf{M}_1$ for $i=1,\cdots,\frac{m+1}2$.
\item $x_1-x_2,y_2-y_1$ in $\mathbf{S}_{m+2+2i}$ are glued to $f,f$ in $\mathbf{M}_1$ for $i=1,\cdots,\frac{m-1}2$.
\item $e-x_2,x_1,e-y_1,y_2$ in $\mathbf{S}_{m+2-2i}$ are glued to $f-x_{2i},y_{2i},x_{2i-1},y_{2i-1}$ in $\mathbf{M}_0$ for $i=1,\cdots,\frac{m+1}2$.
\item $e-x_1,x_2,e-y_2,y_1$ in $\mathbf{S}_{m+2+2i}$ are glued to $f-x_{2i+1},y_{2i+1},x_{2i},y_{2i}$ in $\mathbf{M}_0$ for $i=1,\cdots,\frac{m-1}2$.
\item $e,e$ in $\mathbf{S}_{m+1+2i}$ are glued to $x_{2i}-y_{2i},f-x_{2i-1}-y_{2i-1}$ in $\mathbf{M}_0$ for $i=1,\cdots,\frac{m+1}2$.
\item $e,e$ in $\mathbf{S}_{m+1-2i}$ are glued to $x_{2i+1}-y_{2i+1},f-x_{2i}-y_{2i}$ in $\mathbf{M}_0$ for $i=1,\cdots,\frac{m-1}2$.
\item $x_2-x_1$ in $\mathbf{P}_{0}$ is glued to $f$ in $\mathbf{M}_1$.
\item $f-x_2,x_1$ in $\mathbf{P}_{0}$ is glued to $x_{m+1},y_{m+1}$ in $\mathbf{M}_0$.
\item $f$ in $\mathbf{P}_{1}$ is glued to $f-x_{m+1}-y_{m+1}$ in $\mathbf{M}_0$.
\eit
For $m$ even, the geometry can be written as
\be

\ee
along with the following gluing rules:
\bit
\item $e-x_1-x_2$ in $\mathbf{S}_{m+1}$ is glued to $f$ in $\mathbf{N}_0$.
\item $x_i-x_{i+1}$ in $\mathbf{S}_{m+1}$ is glued to $f$ in $\mathbf{N}_{i}$ for $i=1,\cdots,7$.
\item $x_7,x_8$ in $\mathbf{S}_{m+1}$ are glued to $f-x,y$ in $\mathbf{N}_8$.
\item $e$ in $\mathbf{S}_{m+2}$ is glued to $x-y$ in $\mathbf{N}_8$.
\item $x_1,x_2$ in $\mathbf{S}_{2m+2}$ are glued to $x_1,x_2$ in $\mathbf{P}_0$.
\item $e-x_1-x_2$ in $\mathbf{S}_{2m+2}$ is glued to $f$ in $\mathbf{P}_1$.
\item $e$ in $\mathbf{S}_{1}$ is glued to $f-x_1-x_2$ in $\mathbf{P}_0$.
\item $x_1-x_2$ in $\mathbf{S}_{m+2}$ is glued to $f$ in $\mathbf{M}_1$.
\item $e-x_1,x_2$ in $\mathbf{S}_{m+2}$ are glued to $f-x_{1},y_{1}$ in $\mathbf{M}_0$.
\item $e$ in $\mathbf{S}_{m+1}$ is glued to $x_{1}-y_{1}$ in $\mathbf{M}_0$.
\item $x_2-x_1,y_1-y_2$ in $\mathbf{S}_{m+2-2i}$ are glued to $f,f$ in $\mathbf{M}_1$ for $i=1,\cdots,\frac{m}2$.
\item $x_1-x_2,y_2-y_1$ in $\mathbf{S}_{m+2+2i}$ are glued to $f,f$ in $\mathbf{M}_1$ for $i=1,\cdots,\frac{m}2$.
\item $e-x_2,x_1,e-y_1,y_2$ in $\mathbf{S}_{m+2-2i}$ are glued to $f-x_{2i},y_{2i},x_{2i-1},y_{2i-1}$ in $\mathbf{M}_0$ for $i=1,\cdots,\frac{m}2$.
\item $e-x_1,x_2,e-y_2,y_1$ in $\mathbf{S}_{m+2+2i}$ are glued to $f-x_{2i+1},y_{2i+1},x_{2i},y_{2i}$ in $\mathbf{M}_0$ for $i=1,\cdots,\frac{m}2$.
\item $e,e$ in $\mathbf{S}_{m+1+2i}$ are glued to $x_{2i}-y_{2i},f-x_{2i-1}-y_{2i-1}$ in $\mathbf{M}_0$ for $i=1,\cdots,\frac{m}2$.
\item $e,e$ in $\mathbf{S}_{m+1-2i}$ are glued to $x_{2i+1}-y_{2i+1},f-x_{2i}-y_{2i}$ in $\mathbf{M}_0$ for $i=1,\cdots,\frac{m}2$.
\item $x_1-x_2$ in $\mathbf{P}_{0}$ is glued to $f$ in $\mathbf{M}_1$.
\item $f-x_1,x_2$ in $\mathbf{P}_{0}$ is glued to $x_{m+1},y_{m+1}$ in $\mathbf{M}_0$.
\item $f$ in $\mathbf{P}_{1}$ is glued to $f-x_{m+1}-y_{m+1}$ in $\mathbf{M}_0$.
\eit
The theories in (\ref{t12}) are produced by integrating out $x_i$ living in $\mathbf{S}_{m+1}$, the theories in (\ref{t13}) are produced by integrating out $f-x_1$ before integrating out remaining $x_i$ living in $\mathbf{S}_{m+1}$, and the theories in (\ref{t14}) are produced by integrating out $f-x_1$, $f-x_2$, $x_8$ (in that order) before integrating out remaining $x_i$ living in $\mathbf{S}_{m+1}$.

\subsubsection*{\ubf{Derivation of (\ref{t15})}:}
These theories can be produced by using the KK theory
\be

\ee
for which the corresponding $6d$ SCFT has an $\fe_7\oplus\su(2)^2$ flavor symmetry. For $m$ odd, we write the geometry as
\be
%
\ee
where one of the $\su(2)$ factors in the flavor symmetry is represented in a non-affine form (via surface $\mathbf{M}_1$) in order to simplify the presentation. This lack of information does not influence the computation of flavor symmetry for $5d$ SCFTs appearing in (\ref{t15}) because of the following reason:\\
In the affinized form, this $\su(2)$ flavor symmetry of the $6d$ SCFT appears as two non-compact surfaces $\mathbf{M}_0$, $\mathbf{M}_1$ with intersection matrix being the Cartan matrix for $\su(2)^{(1)}$. After an RG flow to a $5d$ SCFT, either one of these surfaces or both of these must be integrated out since the flavor symmetry of a  $5d$ SCFT can not have an affine Lie algebra as a factor. We will see that this RG flow does not integrate out $\mathbf{M}_1$, thus $\mathbf{M}_0$ must have been integrated out.

The gluing rules for the above geometry are:
\bit
\item $h-x_1-x_2-x_3-x_4$ in $\mathbf{S}_{m+1}$ is glued to $f$ in $\mathbf{N}_7$.
\item $x_i-x_{i+1}$ in $\mathbf{S}_{m+1}$ is glued to $f$ in $\mathbf{N}_{i-1}$ for $i=1,\cdots,7$.
\item $y_1,y_2$ in $\mathbf{S}_{m+2}$ are glued to $x_1,x_2$ in $\mathbf{P}_0$.
\item $e-y_1-y_2$ in $\mathbf{S}_{m+2}$ is glued to $f$ in $\mathbf{P}_1$.
\item $e$ in $\mathbf{S}_{m}$ is glued to $f-x_1-x_2$ in $\mathbf{P}_0$.
\item $x_1-x_2,y_2-y_1$ in $\mathbf{S}_{m+2i}$ are glued to $f,f$ in $\mathbf{M}_1$ for $i=1,\cdots,\frac{m-1}2$.
\item $e-x_1-x_2,x_1-x_2,y_2-y_1$ in $\mathbf{S}_{2m+1}$ are glued to $f,f,f$ in $\mathbf{M}_1$.
\item $x_2-x_1,y_1-y_2$ in $\mathbf{S}_{m+1-2i}$ are glued to $f,f$ in $\mathbf{M}_1$ for $i=1,\cdots,\frac{m-1}2$.
\item $x_2-x_1$ in $\mathbf{P}_{0}$ is glued to $f$ in $\mathbf{M}_1$.
\eit
For even $m$, we write the geometry as
\be

\ee
along with the following gluing rules
\bit
\item $h-x_1-x_2-x_3-x_4$ in $\mathbf{S}_{m+1}$ is glued to $f$ in $\mathbf{N}_7$.
\item $x_i-x_{i+1}$ in $\mathbf{S}_{m+1}$ is glued to $f$ in $\mathbf{N}_{i-1}$ for $i=1,\cdots,7$.
\item $y_1,y_2$ in $\mathbf{S}_{m}$ are glued to $x_1,x_2$ in $\mathbf{P}_0$.
\item $e-y_1-y_2$ in $\mathbf{S}_{m}$ is glued to $f$ in $\mathbf{P}_1$.
\item $e$ in $\mathbf{S}_{m+2}$ is glued to $f-x_1-x_2$ in $\mathbf{P}_0$.
\item $x_2-x_1,y_1-y_2$ in $\mathbf{S}_{m+2-2i}$ are glued to $f,f$ in $\mathbf{M}_1$ for $i=1,\cdots,\frac{m}2$.
\item $x_1-x_2,y_2-y_1$ in $\mathbf{S}_{m+1+2i}$ are glued to $f,f$ in $\mathbf{M}_1$ for $i=1,\cdots,\frac{m-2}2$.
\item $e-x_1-x_2,x_1-x_2,y_2-y_1$ in $\mathbf{S}_{2m+1}$ are glued to $f,f,f$ in $\mathbf{M}_1$.
\item $x_1-x_2$ in $\mathbf{P}_{0}$ is glued to $f$ in $\mathbf{M}_1$.
\eit
After doing some flops, we can write the geometry for $m$ odd as
\be

\ee
along with the following gluing rules
\bit
\item $h-x_1-x_2-x_3$ in $\mathbf{S}_{m+1}$ is glued to $f$ in $\mathbf{N}_7$.
\item $f-x_1$ in $\mathbf{S}_{m+1}$ is glued to $f-x_1$ in $\mathbf{N}_0$.
\item $x_i-x_{i+1}$ in $\mathbf{S}_{m+1}$ is glued to $f$ in $\mathbf{N}_{i}$ for $i=1,\cdots,6$.
\item $f$ in $\mathbf{S}_{m+1+i}$ is glued to $x_i-x_{i+1}$ in $\mathbf{N}_{0}$ for $i=1,\cdots,m-1$.
\item $x_4$ in $\mathbf{S}_{2m+1}$ is glued to $x_m$ in $\mathbf{N}_{0}$.
\item $y_1,y_2$ in $\mathbf{S}_{m+2}$ are glued to $x_1,x_2$ in $\mathbf{P}_0$.
\item $e-y_1-y_2$ in $\mathbf{S}_{m+2}$ is glued to $f$ in $\mathbf{P}_1$.
\item $e$ in $\mathbf{S}_{m}$ is glued to $f-x_1-x_2$ in $\mathbf{P}_0$.
\item $x_1-x_2,y_2-y_1$ in $\mathbf{S}_{m+2i}$ are glued to $f,f$ in $\mathbf{M}_1$ for $i=1,\cdots,\frac{m-1}2$.
\item $e-x_1-x_2,x_1-x_2,y_2-y_1$ in $\mathbf{S}_{2m+1}$ are glued to $f,f,f$ in $\mathbf{M}_1$.
\item $x_2-x_1,y_1-y_2$ in $\mathbf{S}_{m+1-2i}$ are glued to $f,f$ in $\mathbf{M}_1$ for $i=1,\cdots,\frac{m-1}2$.
\item $x_2-x_1$ in $\mathbf{P}_{0}$ is glued to $f$ in $\mathbf{M}_1$.
\eit
Similarly, performing some flops, we can write the geometry for $m$ even as
\be

\ee
along with the following gluing rules
\bit
\item $h-x_1-x_2-x_3$ in $\mathbf{S}_{m+1}$ is glued to $f$ in $\mathbf{N}_7$.
\item $f-x_1$ in $\mathbf{S}_{m+1}$ is glued to $f-x_1$ in $\mathbf{N}_0$.
\item $x_i-x_{i+1}$ in $\mathbf{S}_{m+1}$ is glued to $f$ in $\mathbf{N}_{i}$ for $i=1,\cdots,6$.
\item $f$ in $\mathbf{S}_{m+1+i}$ is glued to $x_i-x_{i+1}$ in $\mathbf{N}_{0}$ for $i=1,\cdots,m-1$.
\item $x_4$ in $\mathbf{S}_{2m+1}$ is glued to $x_m$ in $\mathbf{N}_{0}$.
\item $y_1,y_2$ in $\mathbf{S}_{m}$ are glued to $x_1,x_2$ in $\mathbf{P}_0$.
\item $e-y_1-y_2$ in $\mathbf{S}_{m}$ is glued to $f$ in $\mathbf{P}_1$.
\item $e$ in $\mathbf{S}_{m+2}$ is glued to $f-x_1-x_2$ in $\mathbf{P}_0$.
\item $x_2-x_1,y_1-y_2$ in $\mathbf{S}_{m+2-2i}$ are glued to $f,f$ in $\mathbf{M}_1$ for $i=1,\cdots,\frac{m}2$.
\item $x_1-x_2,y_2-y_1$ in $\mathbf{S}_{m+1+2i}$ are glued to $f,f$ in $\mathbf{M}_1$ for $i=1,\cdots,\frac{m-2}2$.
\item $e-x_1-x_2,x_1-x_2,y_2-y_1$ in $\mathbf{S}_{2m+1}$ are glued to $f,f,f$ in $\mathbf{M}_1$.
\item $x_1-x_2$ in $\mathbf{P}_{0}$ is glued to $f$ in $\mathbf{M}_1$.
\eit
After performing an isomorphism on $\mathbf{S}_{2m+1}$, we can write the geometry for odd $m$ as
\be

\ee
which manifests the $5d$ gauge theory description of the $5d$ KK theory. The following gluing rules are
\bit
\item $h-x_1-x_2-x_3$ in $\mathbf{S}_{m+1}$ is glued to $f$ in $\mathbf{N}_7$.
\item $f-x_1$ in $\mathbf{S}_{m+1}$ is glued to $f-x_1$ in $\mathbf{N}_0$.
\item $x_i-x_{i+1}$ in $\mathbf{S}_{m+1}$ is glued to $f$ in $\mathbf{N}_{i}$ for $i=1,\cdots,6$.
\item $f$ in $\mathbf{S}_{m+1+i}$ is glued to $x_i-x_{i+1}$ in $\mathbf{N}_{0}$ for $i=1,\cdots,m-1$.
\item $e-x_4$ in $\mathbf{S}_{2m+1}$ is glued to $x_m$ in $\mathbf{N}_{0}$.
\item $y_1,y_2$ in $\mathbf{S}_{m+2}$ are glued to $x_1,x_2$ in $\mathbf{P}_0$.
\item $e-y_1-y_2$ in $\mathbf{S}_{m+2}$ is glued to $f$ in $\mathbf{P}_1$.
\item $e$ in $\mathbf{S}_{m}$ is glued to $f-x_1-x_2$ in $\mathbf{P}_0$.
\item $x_1-x_2,y_2-y_1$ in $\mathbf{S}_{m+2i}$ are glued to $f,f$ in $\mathbf{M}_1$ for $i=1,\cdots,\frac{m-1}2$.
\item $f-x_3-x_4,x_2-x_1,y_2-y_1$ in $\mathbf{S}_{2m+1}$ are glued to $f,f,f$ in $\mathbf{M}_1$.
\item $x_2-x_1,y_1-y_2$ in $\mathbf{S}_{m+1-2i}$ are glued to $f,f$ in $\mathbf{M}_1$ for $i=1,\cdots,\frac{m-1}2$.
\item $x_2-x_1$ in $\mathbf{P}_{0}$ is glued to $f$ in $\mathbf{M}_1$.
\eit
Similarly, the geometry for $m$ even takes the form
\be

\ee
along with the following gluing rules
\bit
\item $h-x_1-x_2-x_3$ in $\mathbf{S}_{m+1}$ is glued to $f$ in $\mathbf{N}_7$.
\item $f-x_1$ in $\mathbf{S}_{m+1}$ is glued to $f-x_1$ in $\mathbf{N}_0$.
\item $x_i-x_{i+1}$ in $\mathbf{S}_{m+1}$ is glued to $f$ in $\mathbf{N}_{i}$ for $i=1,\cdots,6$.
\item $f$ in $\mathbf{S}_{m+1+i}$ is glued to $x_i-x_{i+1}$ in $\mathbf{N}_{0}$ for $i=1,\cdots,m-1$.
\item $e-x_4$ in $\mathbf{S}_{2m+1}$ is glued to $x_m$ in $\mathbf{N}_{0}$.
\item $y_1,y_2$ in $\mathbf{S}_{m}$ are glued to $x_1,x_2$ in $\mathbf{P}_0$.
\item $e-y_1-y_2$ in $\mathbf{S}_{m}$ is glued to $f$ in $\mathbf{P}_1$.
\item $e$ in $\mathbf{S}_{m+2}$ is glued to $f-x_1-x_2$ in $\mathbf{P}_0$.
\item $x_2-x_1,y_1-y_2$ in $\mathbf{S}_{m+2-2i}$ are glued to $f,f$ in $\mathbf{M}_1$ for $i=1,\cdots,\frac{m}2$.
\item $x_1-x_2,y_2-y_1$ in $\mathbf{S}_{m+1+2i}$ are glued to $f,f$ in $\mathbf{M}_1$ for $i=1,\cdots,\frac{m-2}2$.
\item $f-x_3-x_4,x_2-x_1,y_2-y_1$ in $\mathbf{S}_{2m+1}$ are glued to $f,f,f$ in $\mathbf{M}_1$.
\item $x_1-x_2$ in $\mathbf{P}_{0}$ is glued to $f$ in $\mathbf{M}_1$.
\eit
The theories in (\ref{t15}) are produced by successively integrating out $x_i$ living in $\mathbf{S}_{m+1}$. As can be seen from the gluing rules, any such RG flow integrates out $\mathbf{P}_0$ for $m$ odd and $\mathbf{P}_1$ for $m$ even, while preserving $\mathbf{M}_1$ in both cases. The effect on $\mathbf{N}_i$ is same for both cases.

\subsubsection*{\ubf{Derivation of (\ref{t16})}:}
These theories can be produced using the KK theory
\be

\ee
for which the corresponding $6d$ SCFT has an $\so(16)\oplus\su(2)$ flavor symmetry. Again, it turns out to be enough to know the coupling of the compact part of the geometry only to non-affinized $\su(2)$, as in the previous case\footnote{The full coupling to $\su(2)^{(1)}$ for this case and the previous case can be found in Part 1 of this series of papers.}. For $m$ even, we write the geometry as
\be
%
\ee
along with the following gluing rules:
\bit
\item $e-x_1-x_2$ in $\mathbf{S}_{m+2}$ is glued to $f$ in $\mathbf{N}_0$.
\item $x_i-x_{i+1}$ in $\mathbf{S}_{m+2}$ is glued to $f$ in $\mathbf{N}_{i}$ for $i=1,\cdots,7$.
\item $x_7,x_8$ in $\mathbf{S}_{m+2}$ are glued to $f-x,y$ in $\mathbf{N}_8$.
\item $e$ in $\mathbf{S}_{m+1}$ is glued to $x-y$ in $\mathbf{N}_8$.
\item $y_1-y_2$ in $\mathbf{S}_{m+2}$ is glued to $f$ in $\mathbf{M}_1$.
\item $x_2-x_1,y_1-y_2$ in $\mathbf{S}_{m+2-2i}$ are glued to $f,f$ in $\mathbf{M}_1$ for $i=1,\cdots,\frac{m}2$.
\item $x_1-x_2,y_2-y_1$ in $\mathbf{S}_{m+2+2i}$ are glued to $f,f$ in $\mathbf{M}_1$ for $i=1,\cdots,\frac{m-2}2$.
\item $e-x_1-x_2,x_1-x_2,y_2-y_1$ in $\mathbf{S}_{2m+2}$ are glued to $f,f,f$ in $\mathbf{M}_1$.
\eit
For $m$ odd, we write the geometry as
\be

\ee
along with the following gluing rules:
\bit
\item $e-x_1-x_2$ in $\mathbf{S}_{m+2}$ is glued to $f$ in $\mathbf{N}_0$.
\item $x_i-x_{i+1}$ in $\mathbf{S}_{m+2}$ is glued to $f$ in $\mathbf{N}_{i}$ for $i=1,\cdots,7$.
\item $x_7,x_8$ in $\mathbf{S}_{m+2}$ are glued to $f-x,y$ in $\mathbf{N}_8$.
\item $e$ in $\mathbf{S}_{m+1}$ is glued to $x-y$ in $\mathbf{N}_8$.
\item $x_2-x_1$ in $\mathbf{S}_{m+1}$ is glued to $f$ in $\mathbf{M}_1$.
\item $x_2-x_1,y_1-y_2$ in $\mathbf{S}_{m+1-2i}$ are glued to $f,f$ in $\mathbf{M}_1$ for $i=1,\cdots,\frac{m-1}2$.
\item $x_1-x_2,y_2-y_1$ in $\mathbf{S}_{m+1+2i}$ are glued to $f,f$ in $\mathbf{M}_1$ for $i=1,\cdots,\frac{m-1}2$.
\item $e-x_1-x_2,x_1-x_2,y_2-y_1$ in $\mathbf{S}_{2m+2}$ are glued to $f,f,f$ in $\mathbf{M}_1$.
\eit
By performing similar manipulations as for the case of (\ref{t15}), we can rewrite the above geometry for $m$ even as
\be

\ee
along with the following gluing rules:
\bit
\item $e-x_1$ in $\mathbf{S}_{m+2}$ is glued to $f$ in $\mathbf{N}_0$.
\item $f-x_1$ in $\mathbf{S}_{m+2}$ is glued to $f-x_1$ in $\mathbf{N}_1$.
\item $x_i-x_{i+1}$ in $\mathbf{S}_{m+2}$ is glued to $f$ in $\mathbf{N}_{i+1}$ for $i=1,\cdots,6$.
\item $x_6,x_7$ in $\mathbf{S}_{m+2}$ are glued to $f-x,y$ in $\mathbf{N}_8$.
\item $e$ in $\mathbf{S}_{m+1}$ is glued to $x-y$ in $\mathbf{N}_8$.
\item $f$ in $\mathbf{S}_{m+2+i}$ is glued to $x_i-x_{i+1}$ in $\mathbf{N}_{1}$ for $i=1,\cdots,m-1$.
\item $e-x_4$ in $\mathbf{S}_{2m+2}$ is glued to $x_m$ in $\mathbf{N}_{1}$.
\item $y_1-y_2$ in $\mathbf{S}_{m+2}$ is glued to $f$ in $\mathbf{M}_1$.
\item $x_2-x_1,y_1-y_2$ in $\mathbf{S}_{m+2-2i}$ are glued to $f,f$ in $\mathbf{M}_1$ for $i=1,\cdots,\frac{m}2$.
\item $x_1-x_2,y_2-y_1$ in $\mathbf{S}_{m+2+2i}$ are glued to $f,f$ in $\mathbf{M}_1$ for $i=1,\cdots,\frac{m-2}2$.
\item $f-x_3-x_4,x_2-x_1,y_2-y_1$ in $\mathbf{S}_{2m+2}$ are glued to $f,f,f$ in $\mathbf{M}_1$.
\eit
and the geometry for odd $m$ as
\be

\ee
along with the following gluing rules:
\bit
\item $e-x_1$ in $\mathbf{S}_{m+2}$ is glued to $f$ in $\mathbf{N}_0$.
\item $f-x_1$ in $\mathbf{S}_{m+2}$ is glued to $f-x_1$ in $\mathbf{N}_1$.
\item $x_i-x_{i+1}$ in $\mathbf{S}_{m+2}$ is glued to $f$ in $\mathbf{N}_{i+1}$ for $i=1,\cdots,6$.
\item $x_6,x_7$ in $\mathbf{S}_{m+2}$ are glued to $f-x,y$ in $\mathbf{N}_8$.
\item $e$ in $\mathbf{S}_{m+1}$ is glued to $x-y$ in $\mathbf{N}_8$.
\item $f$ in $\mathbf{S}_{m+2+i}$ is glued to $x_i-x_{i+1}$ in $\mathbf{N}_{1}$ for $i=1,\cdots,m-1$.
\item $e-x_4$ in $\mathbf{S}_{2m+2}$ is glued to $x_m$ in $\mathbf{N}_{1}$.
\item $x_2-x_1$ in $\mathbf{S}_{m+1}$ is glued to $f$ in $\mathbf{M}_1$.
\item $x_2-x_1,y_1-y_2$ in $\mathbf{S}_{m+1-2i}$ are glued to $f,f$ in $\mathbf{M}_1$ for $i=1,\cdots,\frac{m-1}2$.
\item $x_1-x_2,y_2-y_1$ in $\mathbf{S}_{m+1+2i}$ are glued to $f,f$ in $\mathbf{M}_1$ for $i=1,\cdots,\frac{m-1}2$.
\item $f-x_3-x_4,x_2-x_1,y_2-y_1$ in $\mathbf{S}_{2m+2}$ are glued to $f,f,f$ in $\mathbf{M}_1$.
\eit
The theories in (\ref{t16}) are obtained by successively integrating out $x_i$ living in $\mathbf{S}_{m+2}$ from the above two geometries.

\subsubsection*{\ubf{Derivation of (\ref{t17}) and (\ref{t18})}:}
These theories can be constructed from the KK theory
\be

\ee
The matter content is a bifundamental along with $m$ fundamentals charged under each $\su(m)$. For $m\ge3$, there is a $\u(1)$ symmetry rotating the bifundamental and an $\su(m)^2$ symmetry rotating the two sets of fundamentals. After twisting, the $\u(1)$ survives, while the two flavor $\su(m)$s are identified with each other. Thus, for $m\ge3$, we expect to be able to couple the compact part of the geometry to non-compact $\P^1$ fibered surfaces whose intersection matrix is the Cartan matrix of $\su(m)^{(1)}$.

For $m\ge3$ and $m=2n$, the geometry can be written as
\be
%
\ee
For $m\ge3$ and $m=2n+1$, the geometry can be written as
\be
%
\ee
In both the cases, the gluing rules are as follows:
\bit
\item $z,y_{m-1}$ in $\mathbf{S}_{m}$ are glued to $f-x_1,x_m$ in $\mathbf{N}_0$.
\item $e-z-y_{1}$ in $\mathbf{S}_{m}$ is glued to $f$ in $\mathbf{N}_{m-1}$.
\item $y_{i}-y_{i+1}$ in $\mathbf{S}_{m}$ is glued to $f$ in $\mathbf{N}_{m-1-i}$ for $i=1,\cdots,m-2$.
\item $e$ in $\mathbf{S}_{1}$ is glued to $x_1-x_{2}$ in $\mathbf{N}_{0}$.
\item $f$ in $\mathbf{S}_{i}$ is glued to $x_i-x_{i+1}$ in $\mathbf{N}_{0}$ for $i=2,\cdots,m-1$.
\eit
The theories in (\ref{t17}) for $m\ge3$ are produced by successively integrating out $y_i$ living in $\mathbf{S}_{m}$. To produced the theories in (\ref{t18}), we have to first integrate out $f-y_1$ living in $\mathbf{S}_{m}$, which integrates out $\mathbf{N}_{m-2}$. Then, we successively integrate out other $y_i$ living in $\mathbf{S}_{m}$.

\subsubsection*{\ubf{Derivation of (\ref{t19})}:}
These can be produced using the $5d$ KK theory
\be

\ee
We expect to be able to couple the compact part of the geometry to $\P^1$ fibered non-compact surfaces with intersection matrix being the Cartan matrix for $\su(2m)^{(2)}$. For $m=2n$, the geometry is
\be
%
\ee
and, for $m=2n+1$, the geometry is
\be
%
\ee
The theories in (\ref{t19}) are obtained by successively integrating out $x_i$ living in the top-most compact surface.

\subsection{Rank 1}
\subsubsection*{\ubf{Derivation of (\ref{t20}) and (\ref{t21})}:}
These can be produced by integrating out fundamentals from the KK theory
\be
%
\ee
where the corresponding $6d$ SCFT has an $\fe_8$ flavor symmetry. The geometry for the KK theory is
\be
%
\ee
Removing fundamentals corresponds to successively integrating out $x_i$ from the compact surface. This leads to the enhanced flavor symmetries shown in (\ref{t20}).

The geometry from $\su(2)+\F$ is found to be
\be\label{g8}
%
\ee
which manifests the non-abelian $\su(2)$ part of the $\u(2)$ flavor symmetry. Notice that the blowup $x$ does not enter in the coupling to $\su(2)$ flavor symmetry, implying that as we integrate out $x$, the $\su(2)$ part of the flavor symmetry survives leading to the geometry
\be
%
\ee
which implies (\ref{t21}), that is pure $5d$ $\su(2)_0$ gauge theory has $\su(2)$ flavor symmetry. On the other hand, if we integrate out $f-x$ from (\ref{g8}), then after flopping this curve we obtain
\be
%
\ee
The integrating out process corresponds to sending the volume of $x$ to infinity in the above geometry, which implies that the $\P^1$ fibration for the non-compact surface is destroyed, thus implying that the flavor symmetry of pure $5d$ $\su(2)_\pi$ gauge theory is $\u(1)$.

\subsection{Rank 2}
\subsubsection*{\ubf{Derivation of (\ref{t22}---\ref{te2})}:}
These theories can be produced by integrating out matter from the following KK theory
\be
%
\ee
carries $12$ fundamental hypers rotated by an $\su(12)$ favor symmetry (the overall $\u(1)$ is anomalous). The $5d$ KK theory
\be\label{e5}
%
\ee
is produced by compactifying the $6d$ SCFT with a twist by charge conjugation symmetry which acts on the flavor $\su(12)$ by an outer automorphism, and thus we expect to couple the geometry corresponding to (\ref{e5}) to a collection of non-compact surfaces with intersection matrix $\su(12)^{(2)}$. According to the gluing rules presented in previous sections, the combined geometry (after some flops) can be written as
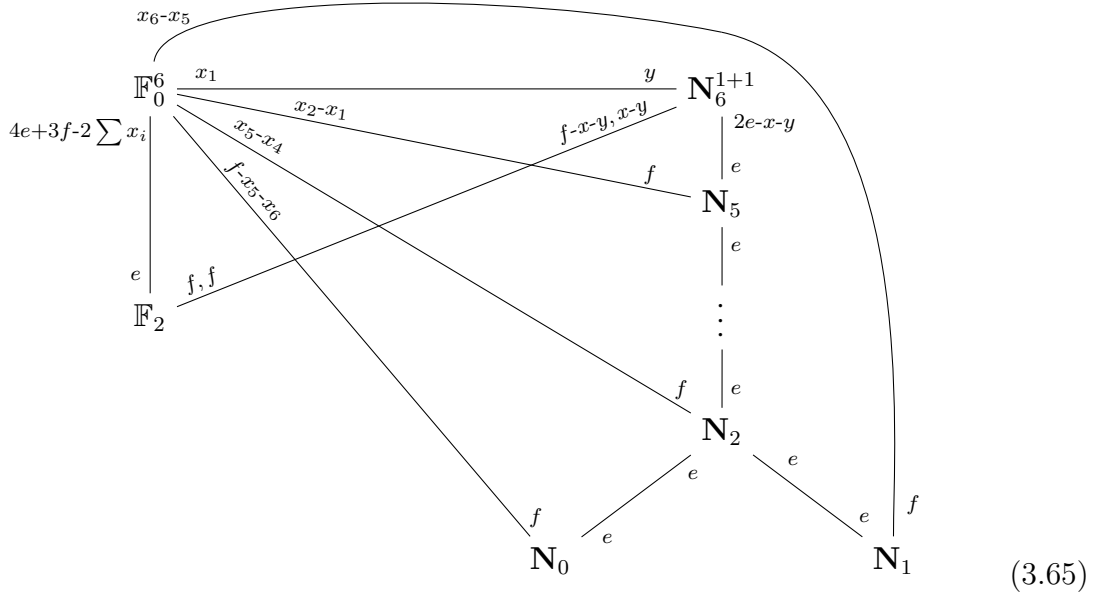
\be
%
\ee
The $\fg_2$ description is obtained after applying $\cS$ on the top-most compact surface
\be
%
\ee
To remove the first fundamental, we have to first flop $f-x_6$ living in the top-most compact surface to obtain
\be
%
\ee
and now we integrate out $f-x$ living in the bottom-most surface which destroys the $\P^1$ fibration of $\mathbf{N}_1$. As a result, we find that $\fg_2+5\F$ has an enhanced $\sp(6)$ flavor symmetry. In a similar fashion, one can successively integrate out the curves $f-x_i$ living in the top-most compact surface to obtain the flavor symmetry for other theories mentioned in (\ref{t22}) and (\ref{t23}).

To derive (\ref{te2}), we write the geometry for $\fg_2+\F$ in the $\su(3)$ frame as follows
\be
%
\ee
Pure $\su(3)_6$ is produced by integrating out $f-x$ living in the top compact surface which preserves $\mathbf{N}_6$, thus leaving an $\su(2)$ flavor symmetry as claimed in (\ref{te2}).

\subsubsection*{\ubf{Derivation of (\ref{t24})}:}
Let us start with the geometry for $\fg_2+3\F$ as derived from the above analysis
\be
%
\ee
We can express this as the geometry for $\sp(2)+2\L^2+\F$ by first applying $\cI_1$ on the top-most compact surface using the blowup $x_3$, and then applying $\cS$ on the top-most compact surface. After performing these isomorphisms, the geometry is written as
\be
%
\ee
Now $\sp(2)_0+2\L^2$ is obtained by integrating out the curve $f-y$ living in the top-most compact surface, which does not intersect any of the $\mathbf{N}_i$. Thus integrating it out does not change the non-abelian part of the flavor symmetry and we recover the result (\ref{t24}).

\subsubsection*{\ubf{Derivation of (\ref{te3})}:}
This can be produced using the KK theory
\be
%
\ee
The matter content is a bifundamental along with $2$ fundamentals charged under each $\su(2)$. There is an $\su(2)$ flavor symmetry rotating the bifundamental and an $\su(2)^2$ flavor symmetry rotating the two sets of fundamentals. After twisting, the $\su(2)$ associated to bifundamental survives, while the other two flavor $\su(2)$s are identified with each other. Thus, we expect to be able to couple the compact part of the geometry to non-compact $\P^1$ fibered surfaces whose intersection matrix is the Cartan matrix of $\su(2)^{(1)}\oplus\su(2)^{(1)}$. Indeed, the geometry can be written as
\be

\ee
The theory in (\ref{te3}) is produced by integrating out $f-x$ living in the top compact surface. This integrates out $\mathbf{N}_0$ and $\mathbf{M}_1$ leaving an $\su(2)^2$ flavor symmetry.

\subsection{Rank 3}
\subsubsection*{\ubf{Derivation of (\ref{t25}) and (\ref{t26})}:}
These theories can be obtained by integrating out matter from the case $m=3$ of (\ref{e6}), but the flavor symmetry of the corresponding $6d$ SCFT is $\su(12)$ instead of $\u(m+8)=\u(11)$. The geometry for $\su(4)_0+\L^2+10\F$ coupled to $\P^1$ fibered non-compact surfaces corresponding to $\su(12)^{(1)}$ is
\be

\ee
The theories (\ref{t25}) are produced by successively integrating out $f-x_i$ living in the top-most compact surface. The theories (\ref{t26}) are produced by first integrating out $f-x$ living in the bottom-most compact surface and then successively integrating out $f-x_i$ living in the top-most compact surface.

\subsubsection*{\ubf{Derivation of (\ref{t27}---\ref{t31})}:}
These can be produced from the KK theory
\be
%
\ee
The $6d$ SCFT has an $\fe_7\oplus\so(7)$ flavor symmetry where the $\fe_7$ part is the flavor symmetry associated to the $\sp(0)$ node and $\so(7)$ is the flavor symmetry associated to the $\su(2)$ node\footnote{Naively one might think that there is an $\so(8)$ flavor symmetry rotating the 4 fundamental hypers charged under $\su(2)$. However, it is known that there is a reduction in the rank of the flavor symmetry and the flavor symmetry is infact $\so(7)$ with the 4 hypers transforming in the strictly-real spinor representation of $\so(7)$.}. The geometry for the KK theory is then found to be
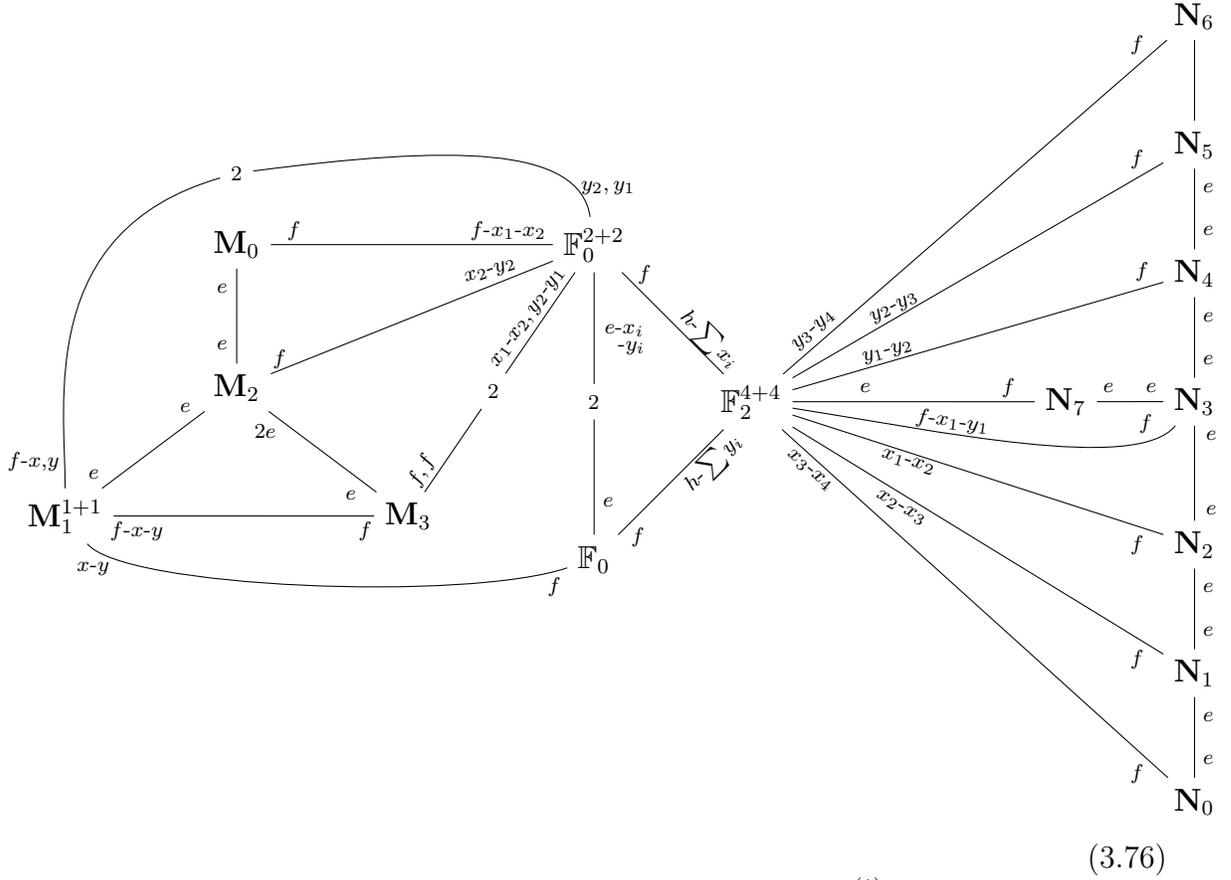
\be

\ee
where $\mathbf{N}_i$ parametrize non-compact surfaces corresponding to $\fe_7^{(1)}$ and $\mathbf{M}_i$ parametrize non-compact surfaces corresponding to $\so(7)^{(1)}$. $\su(4)_0+2\L^2+8\F$ is obtained by applying $\cS$ on the top-most and bottom-most compact surfaces to obtain
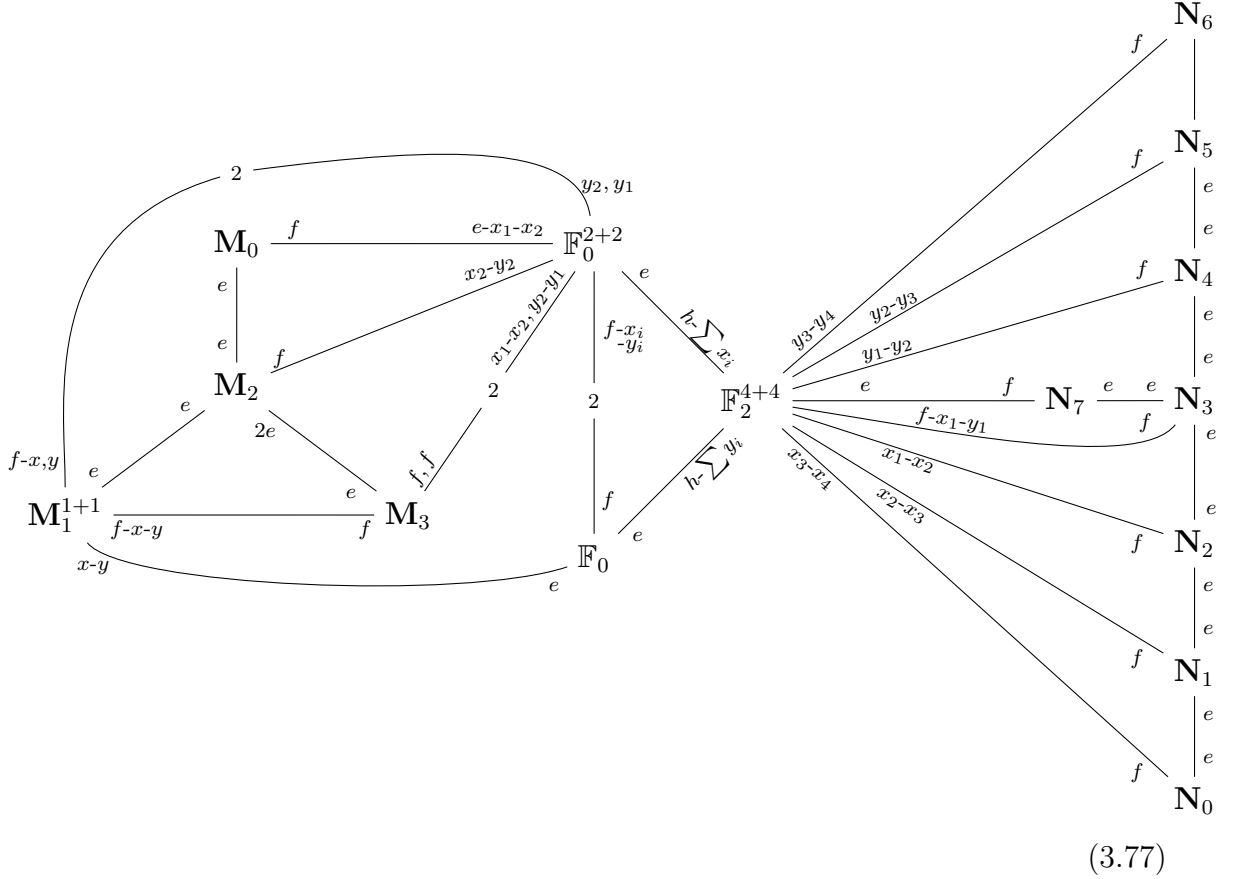
\be
%
\ee
We can then produce $\su(4)_\half+2\L^2+7\F$ by integrating out $y_4$ living in the middle compact surface, which removes $\mathbf{N}_6$ and $\mathbf{M}_1$ implying that the flavor symmetry is $\fe_7\oplus\so(7)$. Removing other $y_i$ living in the middle compact surface we reach $\su(4)_2+2\L^2+4\F$. To go beyond this point and obtain other theories in (\ref{t27}), we need to successively integrate out the curves $f-x_i$ living in the middle compact surface. The reader can verify that these processes lead to the flavor symmetry claimed in (\ref{t27}). 

To obtain theories in (\ref{t28}), we first integrate out $x_4$, which decouples $\mathbf{N}_0$ and $\mathbf{M}_0$, and then successively integrate out $y_i$ living in the middle compact surface until we reach $\su(2)_\frac32+2\L^2+3\F$, from which point onward we integrate out the remaining $f-x_i$ living in the middle compact surface. Similarly, to obtain theories in (\ref{t29}), we first integrate out $x_4,x_3,y_4$ (in that order), which decouples $\mathbf{N}_0$, $\mathbf{N}_1$, $\mathbf{N}_6$, $\mathbf{M}_0$ and $\mathbf{M}_1$, and then successively integrate out $y_i$ living in the middle compact surface until we reach $\su(2)_1+2\L^2+2\F$, from which point onward we integrate out the remaining $f-x_i$ living in the middle compact surface. In a similar fashion, we can also obtain theories in (\ref{t30}) and (\ref{t31}) by integrating out more $x_i$ before we start integrating out $y_i$ and $f-x_i$.

\subsubsection*{\ubf{Derivation of (\ref{t32})}:}
These theories can be produced from the KK theory
\be

\ee
The $\su(1)$ node as an $\sp(1)$ flavor symmetry which is fully gauged by the $\sp(1)$ node. The $\sp(1)$ node carries $10\F$ of $\sp(1)$ but a $\half\F$ is trapped in coupling to $\su(1)$, leaving an $\so(19)$ flavor symmetry for the corresponding $6d$ SCFT. The geometry for the $5d$ KK theory turns out to be
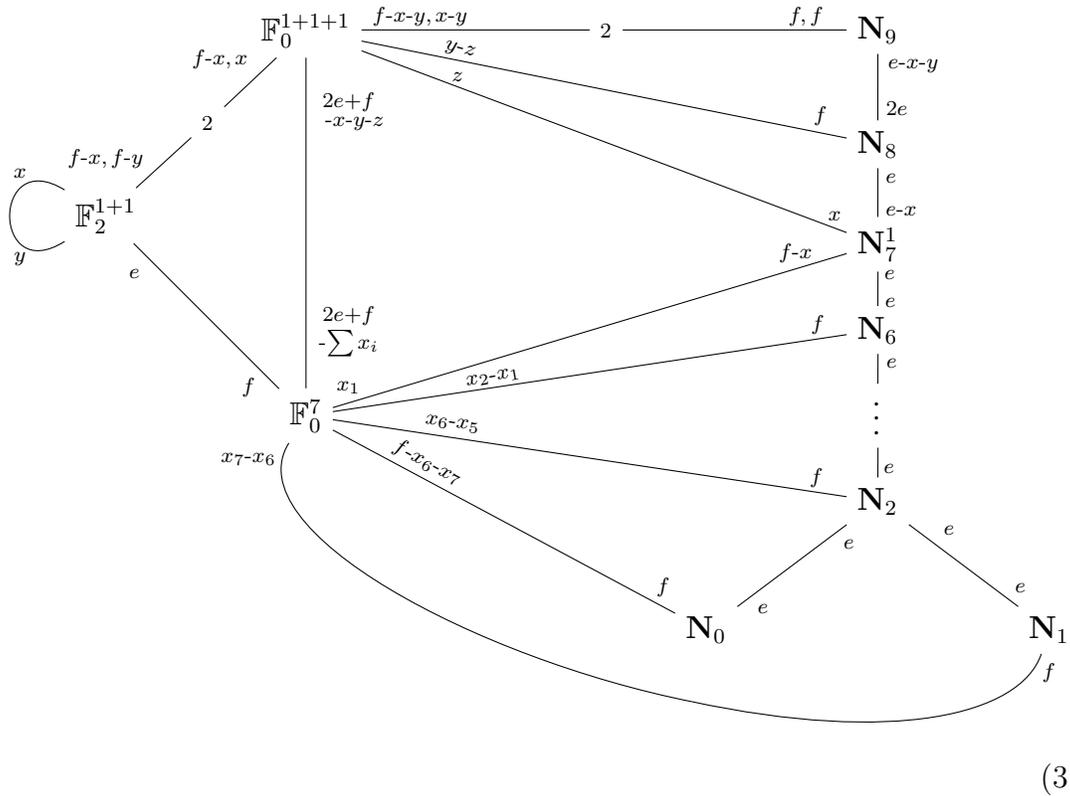
\be
%
\ee
To obtain the $\su(4)_\frac32+2\L^2+7\F$ of the above geometry, we have to first apply $\cS$ on the top-most and bottom-most compact surfaces, and flop $x,y$ living in the middle compact surface, to obtain
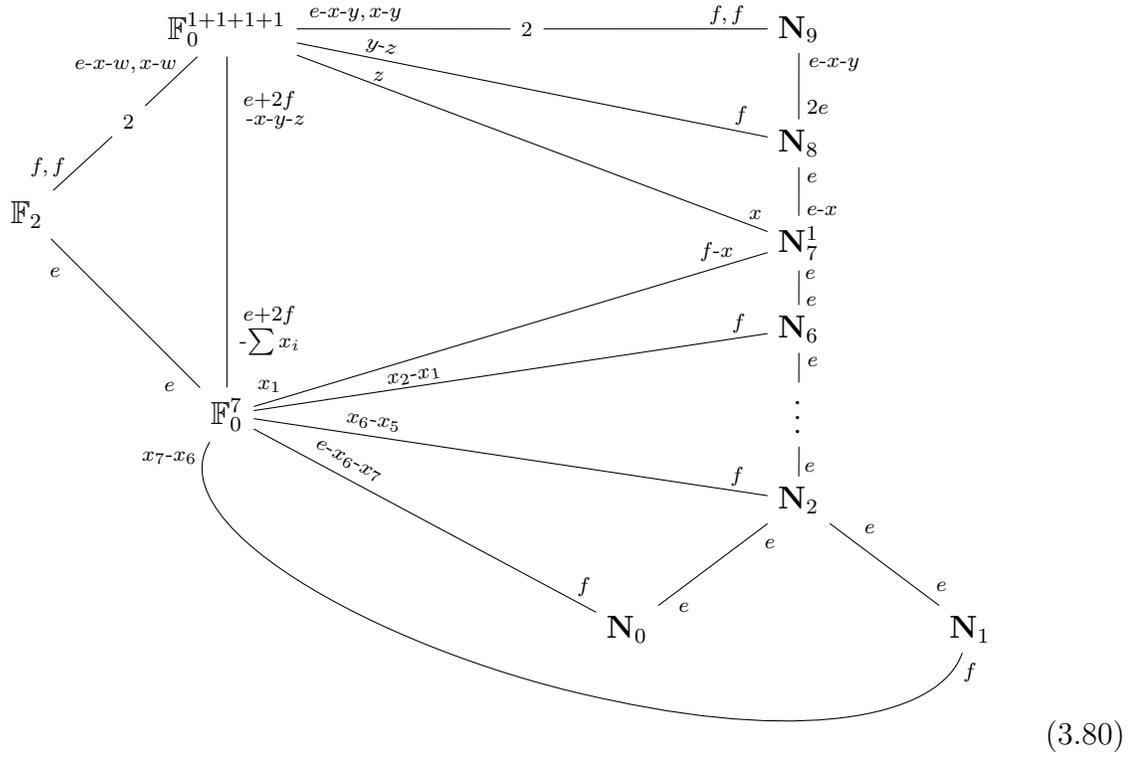
\be
%
\ee
Applying some isomorphisms upon the top-most compact surface, the above geometry can now be written as
\be
%
\ee
which indeed gives rise to the theory $\su(4)_\frac32+2\L^2+7\F$. Now the theories (\ref{t32}) are produced by successively integrating out curves $f-x_i$ living in the bottom-most surface, which leads to the pattern of enhanced flavor symmetries claimed in (\ref{t32}).

When all the fundamentals are integrated out, we obtain an $\so(5)\oplus\u(1)$ flavor symmetry, which is the classical flavor symmetry, not only for $\su(4)_5+2\L^2$, but also for the dual gauge theory $\sp(3)+\half\L^3+\frac52\F$. Thus, the theories obtained by integrating out more fundamentals from $\sp(3)+\half\L^3+\frac52\F$ would have no enhancement of flavor symmetry either.

\subsubsection*{\ubf{Derivation of (\ref{t33}) and (\ref{t34})}:}
For the theories in (\ref{t33}), we use the KK theory
\be

\ee
which allows the coupling of non-compact surfaces comprising an $\fe_6^{(1)}$ as shown in the geometry below
\be
%
\ee
Here we are displaying the compact surfaces in the geometry such that they manifest the $5d$ $\sp(3)$ gauge theory description. To manifest the KK theory description, one should apply $\cS$ on the top-most compact surface. The $5\F$ of $\sp(3)$ originate from the blowups $x_i$ living in the compact surface denoted by $\bF_1^{5+1+1+1}$. Integrating out these blowups leads to the results claimed in (\ref{t33}).

Now, let us consider the geometry when all $5\F$ of $\sp(3)$ have been integrated out
\be
%
\ee
where we manifest the $\su(2)$ non-abelian flavor symmetry of the theory. After performing some flops detailed in \cite{Bhardwaj:2020gyu}, we can write the above geometry as
\be
%
\ee
Applying $\cS$ on the bottom-most compact surface of the above geometry, we obtain a geometry for $\su(4)_\frac{13}2+\F$
\be
%
\ee
Now, we see that the curve $f-x$ living in the bottom-most compact surface of the above geometry does not intersect the non-compact surface $\mathbf{N}_5$. Thus, the process of integrating it out preserves the non-abelian flavor symmetry $\su(2)$, and we obtain the result (\ref{t34}).

\subsubsection*{\ubf{Derivation of (\ref{t35}---\ref{t40})}:}
We start with the $6d$ SCFT
\be
%
\ee
which has $\sp(7)$ flavor symmetry. The untwisted compactification of the above $6d$ SCFT is known to give rise to the $5d$ gauge theory $\so(7)+5\S+2\F$ \cite{Bhardwaj:2020gyu}. The geometry for this $5d$ gauge theory can then be figured out to be
\be\label{g9}
%
\ee
The two fundamentals of $\so(7)$ are encoded differently. One of the fundamentals corresponds to the blowups $x,y$ in the bottom-most compact surface, and the other fundamental corresponds to the blowup $y$ in the middle compact surface. We can integrate out one of the fundamentals by integrating out the curve $f-y$ living in the middle surface. This process integrates out $\mathbf{N}_6$, thus leading to the $\sp(6)\oplus\su(2)$ flavor symmetry claimed in (\ref{t35}).

Another fundamental can be integrated out by integrating out $x,y$ living in the bottom-most compact surface, which integrates out $\mathbf{N}_7$ and $\mathbf{N}_5$ thus verifying the $n=1$ case of (\ref{t36}).

The $5\S$ are encoded in the 5 blowups $x_i$ living in the middle compact surface. They are integrated out by successively integrating out $f-x_i$. Integrating out $f-x_5$ integrates out $\mathbf{N}_4$ thus verifying the $n=2$ case of (\ref{t36}). The reader can check that integrating out further $\S$ leads to a flavor symmetry pattern which shows no enhancement.

Similarly, the reader can also recover the results claimed in (\ref{t37}) and (\ref{t38}) by using the above geometry.

To obtain (\ref{t39}), let us have a look at the geometry for $\so(7)+3\S+2\F$
\be

\ee
Flopping $x,y$ living in the bottom-most compact surface, we get
\be
%
\ee
which can be written, after applying some isomorphisms on the middle compact surface, as
\be
%
\ee
which identifies the above geometry as describing $\su(4)_2+3\L^2+2\F$. We can obtain the $n=1$ case of (\ref{t39}) by integrating out $f-y_2$ living in the middle compact surface, and the $n=2$ case of (\ref{t39}) by further integrating out $y_1$ living in the middle compact surface. The reader can check that this leads to the results claimed in (\ref{t39}).

On the other hand, if we would integrate out $f-y_1$ after integrating out $f-y_2$ from the above geometry, we would obtain $\su(4)_3+3\L^2$ and read the flavor symmetry from the geometry to be $\sp(3)\oplus\u(1)$ which is indeed the classical flavor symmetry.

To obtain (\ref{t40}), we start from the geometry for $\so(7)+4\S+2\F$ obtained from (\ref{g9}) and convert it into a geometry for $\su(4)_\frac32+3\L^2+3\F$ in a similar way as explained above
\be

\ee
from which $\su(4)_0+3\L^2$ is produced by integrating out $f-z$ and $y_i$ living in the middle compact surface. As is clear from the above geometry, these processes integrate out $\mathbf{N}_6$ and $\mathbf{N}_7$ leaving an $\sp(4)$ flavor symmetry as claimed in (\ref{t40}).

\subsubsection*{\ubf{Derivation of (\ref{t41}---\ref{t43})}:}
These theories can be manufactured starting from the KK theory
\be
%
\ee
This theory allows the coupling of non-compact surfaces describing $\fe_6^{(2)}\oplus\su(6)^{(2)}$. This is because the corresponding $6d$ SCFT has an $\fe_6$ flavor symmetry coming from the $\sp(0)$ node and an $\su(6)$ flavor symmetry coming from the $\su(3)$ node as it carries 6 hypers transforming in fundamental of $\su(3)$. A discrete symmetry of the theory can be constructed if a $\Z_2$ outer automorphism acts simultaneously on all these algebras.

According to \cite{Bhardwaj:2020gyu}, this KK theory can be described as $\so(7)+4\S+3\F$ and the geometry can be figured out to be
\be

\ee
The spinors are integrated out by integrating out $y_i$ living in the middle compact surface. The reader can verify that this leads to the results claimed in (\ref{t41}) and (\ref{t42}).

For (\ref{t43}), we can represent the geometry for $\so(7)+3\S+3\F$, after some flops as
\be
%
\ee
Applying $\cS$ on the bottom-most compact surface, we obtain an $\su(4)_\frac52+3\L^2+3\F$ description of the geometry
\be
%
\ee
The theories in (\ref{t43}) are then obtained by integrating out $f-x_i$ living in the bottom-most compact surface.

\subsubsection*{\ubf{Derivation of (\ref{t44})}:}
These theories can be produced from the KK theory
\be
%
\ee
where we have explicitly shown the $\sp(3)^3$ flavor symmetry since the matter content charged under $\so(8)$ is $3\F+3\S+3\C$. A $\Z_3$ outer automorphism of $\so(8)$ cyclically permutes the three kinds of matter contents thus permuting the three $\sp(3)$ flavor symmetries. Consequently, we can represent the KK theory obtained after the twist as
\be
%
\ee
and thus we should use the corresponding gluing rules
\be
%
\ee
to obtain the correct geometry, which can be worked out to be
\be
%
\ee
After applying some isomorphisms on the top-most compact surface, we can rewrite the above geometry as
\be
%
\ee
which manifests the $\sp(3)+\frac12\L^3+\L^2+\frac{5}2\F$ description of the KK theory. The fundamentals can now be integrated out by integrating out the blowups $x_1,y_1$ living in the top-most compact surface. Integrating out $x_1$ integrates out $\mathbf{N}_3$ leaving an $\sp(3)$ flavor symmetry, and further integrating out $y_1$ integrates out $\mathbf{N}_2$ leaving an $\sp(2)$ flavor symmetry.

\subsubsection*{\ubf{Derivation of (\ref{t45})}:}
This theory can be produced by using the KK theory
\be
%
\ee
The non-compact surfaces coupled to the above KK theory comprise an $\su(13)^{(2)}$ and the geometry can be written in the following fashion (after some flops)
\be
%
\ee
which describes $\so(7)+2\S+4\F$ gauge theory. We can remove a spinor by integrating out the blowup $x$ living in the bottom-most compact surface which integrates out $\mathbf{N}_6$ leaving an $\sp(6)$ flavor symmetry.

\subsection{Rank 4}\label{R4d}
\subsubsection*{\ubf{Derivation of (\ref{t46}) and $m=4$ case of (\ref{t5})}:}
These theories can be obtained by integrating out matter from the case $m=4$ of (\ref{e6}), but the flavor symmetry of the corresponding $6d$ SCFT is $\su(12)\oplus\su(2)$ instead of $\u(m+8)=\u(12)$. The geometry is
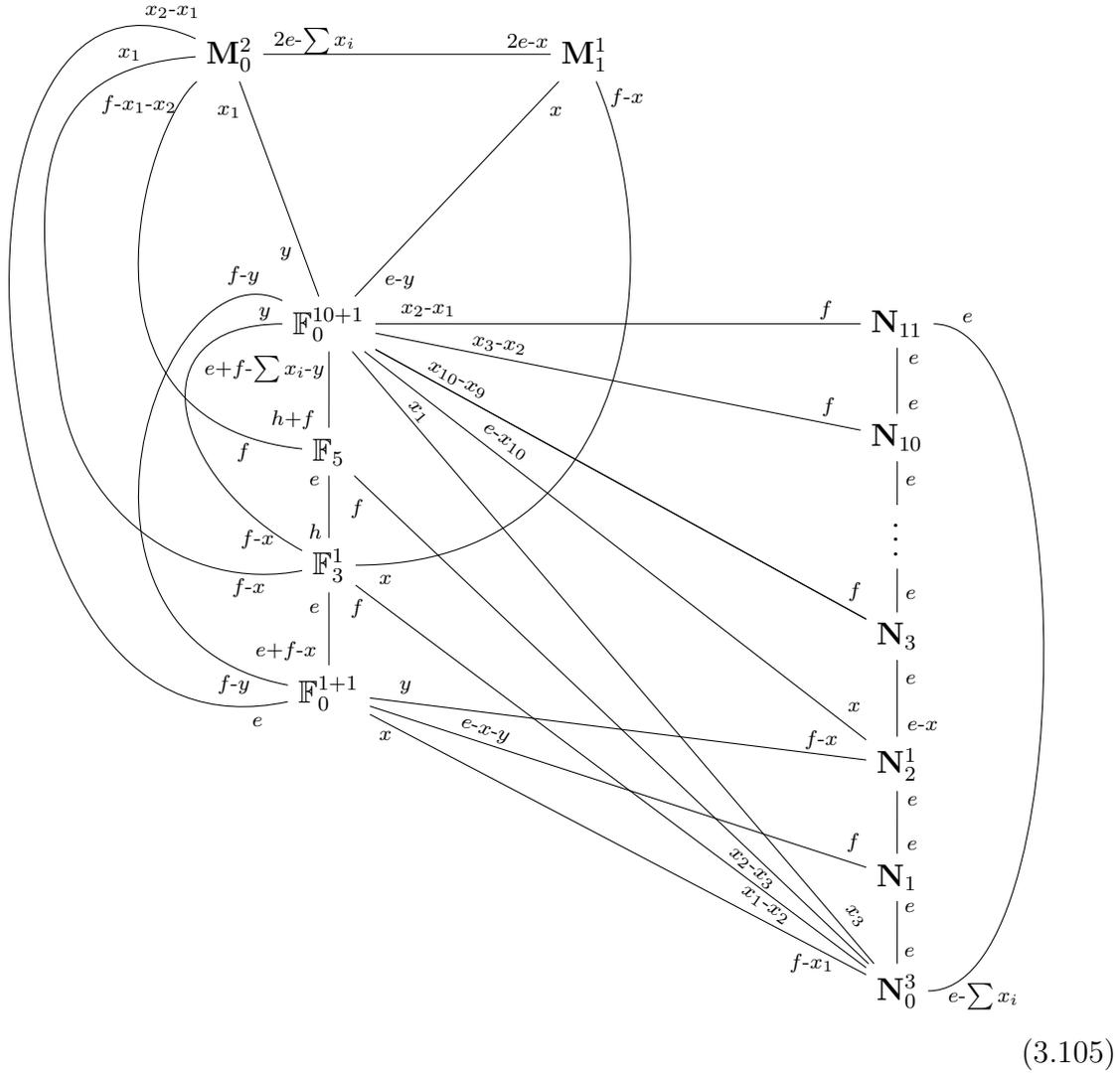
\be
%
\ee
The theories (\ref{t46}) are produced by successively integrating out $f-x_i$ living in the top-most compact surface. $m=4$ case of (\ref{t5}) is produced by first integrating out $x_1$ living in the top-most compact surface and then successively integrating out $f-x_i$ living in the top-most compact surface.

\subsubsection*{\ubf{Derivation of (\ref{t47})}:}
These theories can be produced using the KK theory
\be
%
\ee
The $\sp(0)$ node int he corresponding $6d$ SCFT gives rise to an $\fe_7$ flavor symmetry, and as we have discussed above, if the $\su(1)$ node was absent, then the flavor symmetry corresponding to the $\su(2)$ node would be $\so(7)$, so that the 8 half-hypers in the fundamental of $\su(2)$ transform as a spinor of $\so(7)$. We know that the $\su(1)$ node traps a half-hyper in the fundamental. Correspondingly, $\so(7)$ is broken to $\fg_2$ with the 7 remaining half-hypers transforming in $\F$ of $\fg_2$. The associated geometry can be worked out to be
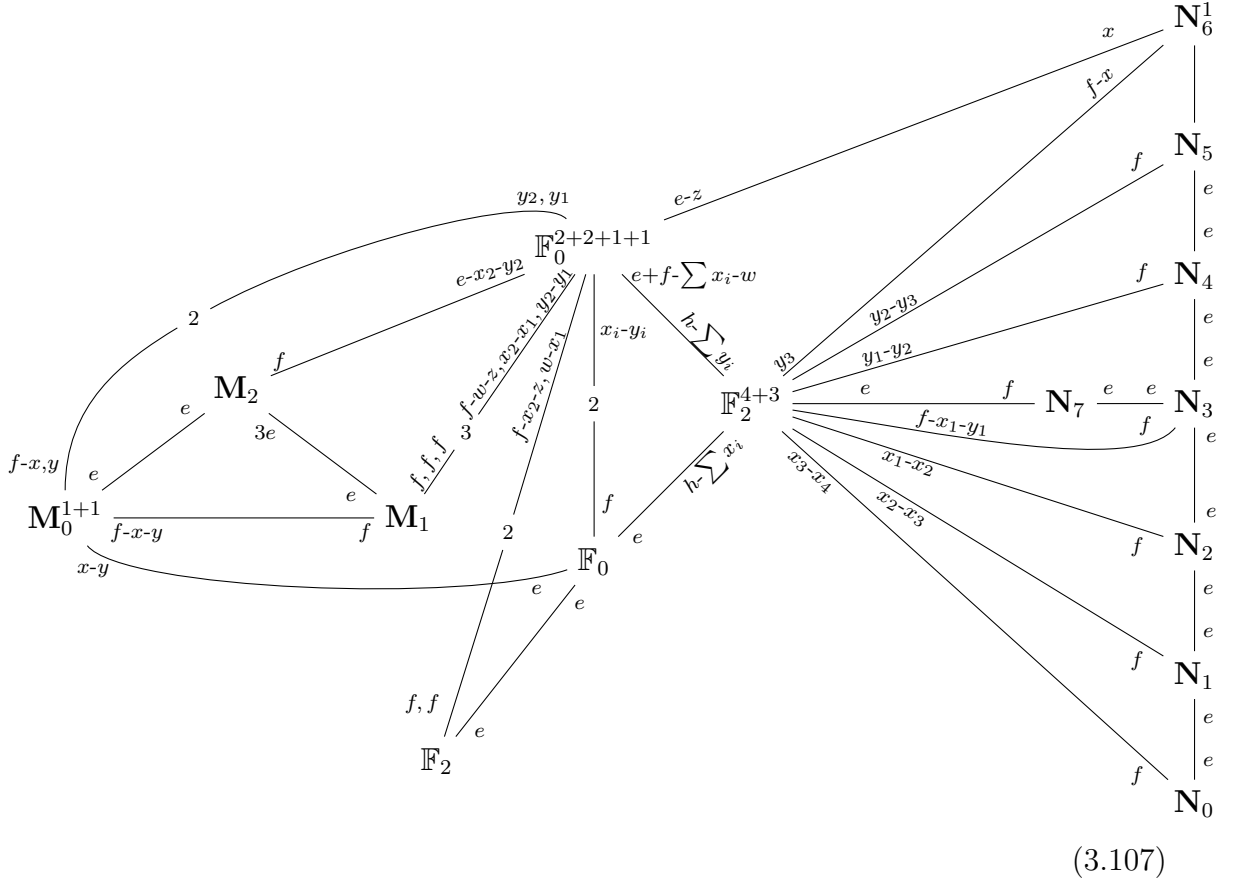
\be
%
\ee
The theories in (\ref{t47}) are produced by integrating out $x_i$ followed by $f-y_i$ living in the compact surface denoted by $\bF_2^{4+3}$ in the above geometry.

\subsubsection*{\ubf{Derivation of (\ref{t48}) and (\ref{t49})}:}
These theories can be produced by starting from the $6d$ SCFT (\ref{e7}), but this time compactifying it with a $\Z_2$ outer automorphism twist instead of the $\Z_3$ twist employed there. Thus, we can represent the KK theory as
\be
%
\ee
The geometry can be worked out to be
\be
%
\ee
where we have manifested the $\so(9)+3\S+3\F$ description. The KK theory description can be manifested by applying $\cS$ on the compact surface $\bF_0$ placed at the second position from the bottom of the diagram. The fundamentals can be integrated out by integrating out the curves $x_i$ living in the surface labeled $\bF_0^{3+3+3}$. The reader can check that this leads to (\ref{t48}). 

To obtain (\ref{t49}), we have to first apply $\cS$ on $\bF_0^{3+3+3}$ in the above geometry to obtain
\be
%
\ee
which manifests the $\su(5)_0+3\L^2+3\F$ description. Then, we integrate out $f-x_1$ living in $\bF_0^{3+3+3}$ to obtain the geometry for $\su(5)_{-\half}+3\L^2+2\F$
\be
%
\ee
The theories in (\ref{t49}) are then obtained by integrating out $x_i$ living in $\bF_1^{2+2+3}$.

\subsubsection*{\ubf{Derivation of (\ref{t50}) and (\ref{t51})}:}
We can manufacture these theories by using the KK theory
\be
%
\ee
whose geometry is
\be
%
\ee
The fundamental of $\so(9)$ can be integrated out by integrating out $x,y$ living in $\bF_2^{1+1}$. This integrates out $\mathbf{M}_0$ and $\mathbf{N}_4$ leaving an $\su(2)\oplus\sp(4)$ flavor symmetry as claimed in (\ref{t50}). After doing some flops and isomorphisms (whose details can be found in \cite{Bhardwaj:2020gyu}) on the geometry associated to $\so(9)+4\S$, we can rewrite it as follows
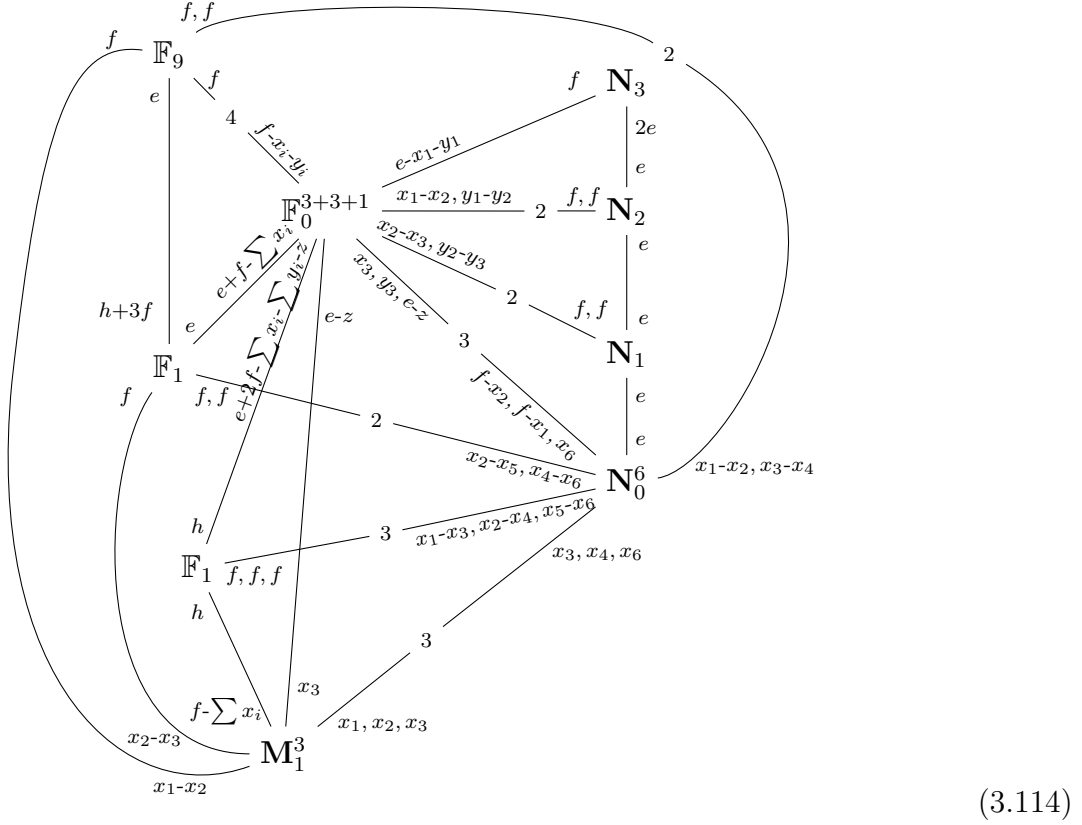
\be
%
\ee
which describes $\su(5)_2+3\L^2+\F$. The $\so(9)$ description can be recovered by applying $\cS$ onto $\bF_0^{3+3+1}$. Now to obtain (\ref{t51}), we have to integrate out $f-z$ living in $\bF_0^{3+3+1}$ which integrates out $\mathbf{N}_3$ and leads to the claimed flavor symmetry.

\subsubsection*{\ubf{Derivation of (\ref{t52})}:}
These can be produced by integrating out matter from the KK theory
\be
%
\ee
whose associated geometry is
\be
%
\ee
The KK theory description is recovered by applying $\cS$ on the surface labeled as $\bF_0$ (without any blowups). The theories in (\ref{t52}) are obtained by successively integrating out $z_i$ living in the top-most compact surface.

\subsubsection*{\ubf{Derivation of (\ref{t53})}:}
These theories can be obtained from the KK theory
\be
%
\ee
whose associated geometry is
\be
%
\ee
where we have made the $\so(9)$ description manifest. To make the KK description manifest, the reader can apply $\cS$ on the surface labeled $\bF_0$. (\ref{t53}) are obtained by integrating out $x_i$ living in the compact surface $\bF_1^5$. 

(\ref{t54}) can be obtained by integrating out a spinor, which corresponds to integrating out half of the blowups blowups living in $\bF_2^{2+2+2+2}$. This process involves too many flops, hence we now turn to the discussion of another KK theory which allows an easier derivation of (\ref{t54}).

\subsubsection*{\ubf{Derivation of (\ref{t54})}:}
To produce the theories $\so(9)+\S+(6-n)\F$, we proceed using the KK theory
\be

\ee
whose associated geometry takes the form
\be
%
\ee
The fundamentals are integrated out by integrating out $x_i$ living in $\bF_1^6$. Integrating out $x_1$, we see that $\mathbf{N}_5$ is integrated out, leading to an $\sp(2)\oplus\sp(5)$ flavor symmetry, as claimed in (\ref{t54}). Subsequently integrating out $x_2$ leads to integrating out $\mathbf{N}_4$ and $\mathbf{N}_6$ leading to an $\sp(4)\oplus\sp(1)\oplus\u(1)$ flavor symmetry, which is just the classical flavor symmetry associated to $\so(9)+\S+4\F$.

\subsubsection*{\ubf{Derivation of (\ref{t55})}:}
These theories can be constructed using the KK theory
\be

\ee
whose geometry is
\be
%
\ee
We can generate theories in (\ref{t55}) by successively integrating out $y_i$ living in the right $\bF_2^{4+4}$. At the first step, this integrates out $\mathbf{N}_0$ and $\mathbf{M}_4$, indeed leading to an $\sp(4)\oplus\ff_4$ flavor symmetry. The reader can easily verify the other results claimed in (\ref{t55}).

\subsubsection*{\ubf{Derivation of (\ref{t56})}:}
We can produce it by integrating out $\C$ from the KK theory
\be
%
\ee
whose associated geometry is
\be
%
\ee
$\C$ can be integrated out by integrating out $x_2$ living in $\bF_0^{2+2}$ which integrates out $\mathbf{N}_0$ leaving an $\sp(7)$ flavor symmetry.

\subsubsection*{\ubf{Derivation of (\ref{t57}) and (\ref{t58})}:}
These can be constructed using the KK theory
\be
%
\ee
whose associated geometry can be written as
\be
%
\ee
$\F$ can be integrated out by integrating out $f-x_4$ living in $\bF_0^4$ which integrates out $\mathbf{N}_3$ and $\mathbf{M}_3$ leaving an $\sp(3)^2\oplus\su(2)^2$ flavor symmetry thus verifying (\ref{t57}). Spinors $\S$ can be integrated by successively integrating out $x_i$ living in $\bF_2^{3+3+1+1}$. At the first step, we integrate out $x_3$, which integrates out $\mathbf{M}_0$ and $\mathbf{N}_4$ leaving an $\sp(4)\oplus\so(9)$ flavor symmetry. At the next step, we further integrate out $x_2$, which further integrates out $\mathbf{M}_1$ leaving an $\sp(4)\oplus\su(4)$ flavor symmetry. This verifies (\ref{t58}).

\subsubsection*{\ubf{Derivation of (\ref{t59})}:}
These theories can be produced using the KK theory
\be

\ee
whose associated geometry can be written as
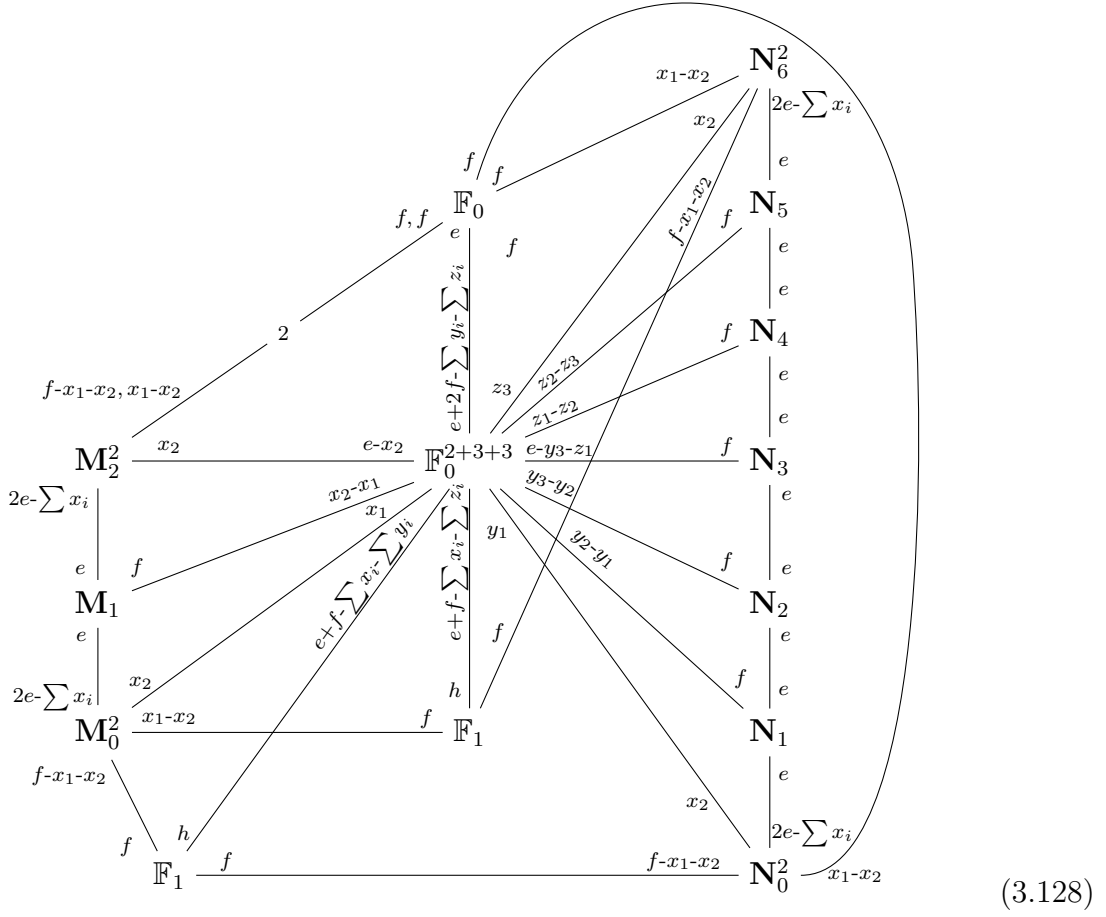
\be
%
\ee
An $\S$ can be removed by integrating out $f-y_3$ living in $\bF_0^{2+3+3}$. This destroys the $\P^1$ fibrations of $\mathbf{N}_2$ and $\mathbf{M}_2$ leaving an $\sp(2)^2\oplus\sp(4)$ flavor symmetry, thus verifying $n=1$ case of (\ref{t59}). The geometry for $\so(8)+2\S+2\C+3\F$ is
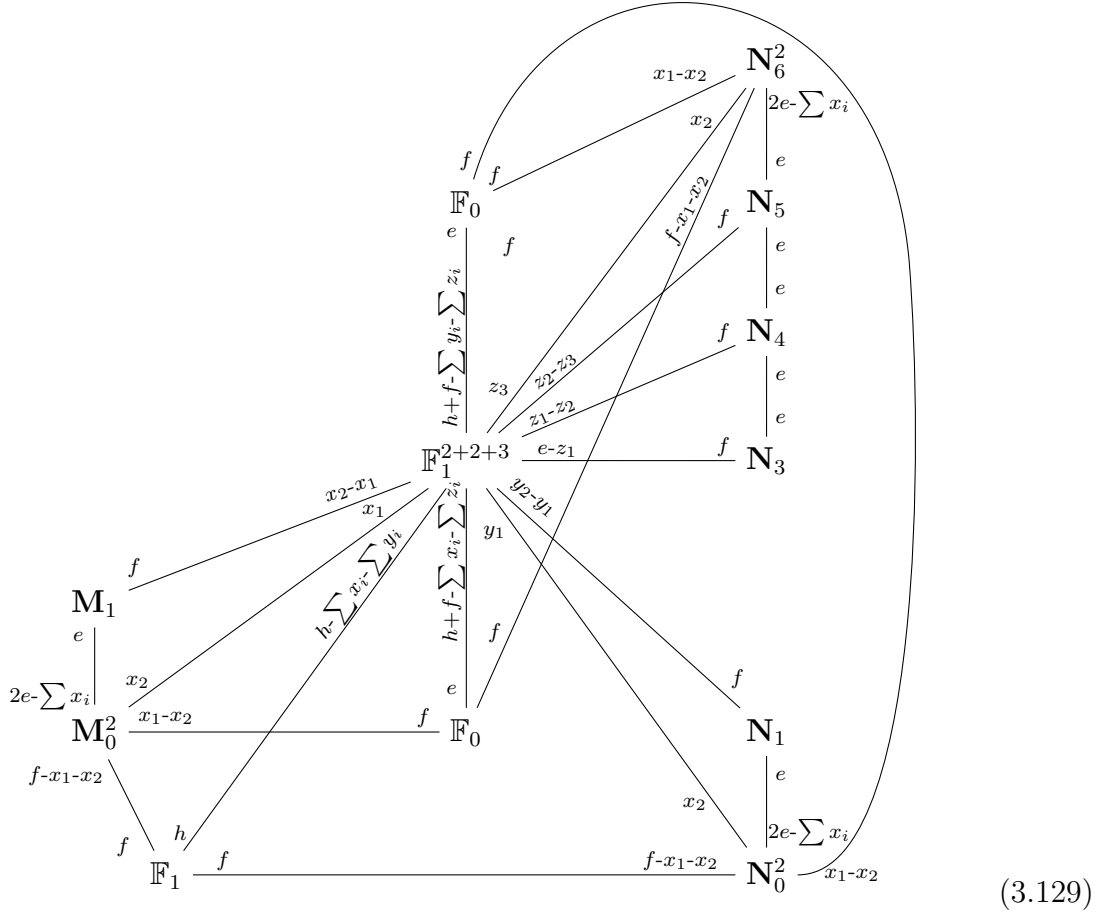
\be
%
\ee
$\F$ can be integrated out by successively integrating out $f-z_i$ living in $\bF_1^{2+2+3}$. Integrating out $f-z_1$ integrates out $\mathbf{N}_4$ leaving $\sp(2)^3\oplus\su(2)$ flavor symmetry, thus verifying $n=2$ case of (\ref{t59}).

\subsection{Rank 5}
\subsubsection*{\ubf{Derivation of (\ref{t60})}:}
These theories can be produced starting from the KK theory $\su(6)_0+\half\L^3+13\F$ which is dual to $\su(6)_0+\L^2+12\F$ \cite{Bhardwaj:2020gyu}. Moreover, this duality holds true when a fundamental is integrated out from both sides leading to
\be
\su(6)_{\frac 12}+\L^2+11\F\:=\:\su(6)_{\frac 12}+\half\L^3+12\F
\ee
The flavor symmetry for the corresponding $5d$ SCFT is already known from (\ref{t4}) to be $\u(13)$, thus recovering the $n=1$ case of (\ref{t60}).

To obtain the $n=2$ case of (\ref{t60}), we need to look at the geometry for 
\be

\ee
in more detail as in following
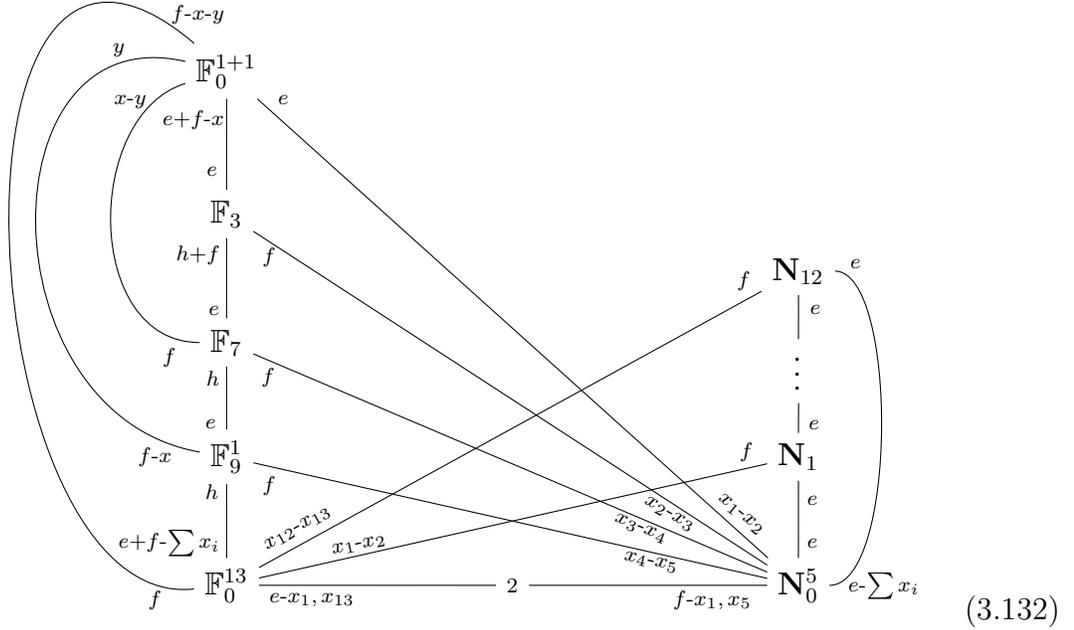
\be
%
\ee
The above geometry manifests the $\su(6)_0+\half\L^3+13\F$ description of the theory. The KK theory description can be manifested by applying $\cS$ on the top-most and the bottom-most compact surfaces. To obtain the $n=2$ case of (\ref{t60}), we need to integrate out $x_{13}$ and $f-x_1$ living in the bottom-most compact surface. $x_{13}$ integrates $\mathbf{N}_{12}$ and $f-x_1$ integrates out $\mathbf{N}_{1}$, thus leaving a $\u(11)\oplus\su(2)$ flavor symmetry as claimed in (\ref{t60}).

\subsubsection*{\ubf{Derivation of (\ref{t61})}:}
For $1\le n\le 8$, these theories are dual to
\be
\su(6)_{3+\frac n2}+\half\L^3+(9-n)\F\:=\:\sp(5)+\L^2+(8-n)\F
\ee
which were already studied in (\ref{t6}). To study the $n=9$ case, we consider the geometry for $\sp(5)+\L^2$
\be

\ee
which is isomorphic to
\be
%
\ee
thus implying that the geometry also constructs $\su(6)_7+\half\L^3+\F$. Integrating out $y$ in top-most compact surface leads to $\su(6)_\frac{15}2+\half\L^3$, and since $y$ does not intersect $\mathbf{N}$, we verify the result quoted in (\ref{t61}).

\subsubsection*{\ubf{Derivation of (\ref{t62}) and (\ref{t63})}:}
These theories can be constructed by using the KK theory
\be
%
\ee
whose associated geometry can be found to be
\be
%
\ee
To manifest the KK description, one should apply $\cS$ on all the compact surfaces having degree zero. The theories (\ref{t62}) are produced by successively integrating out $y_i$ living in $\bF_0^{1+9+1}$. The theories (\ref{t63}) are produced by first integrating out $f-y_1$ living in $\bF_0^{1+9+1}$ and then successively integrating out $y_i$.

\subsubsection*{\ubf{Derivation of (\ref{t64})}:}
These theories are dual to
\be
\su(6)_\frac{3+n}2+2\L^2+(7-n)\F\:=\:\su(6)_\frac{3+n}{2}+\half\L^3+\L^2+(8-n)\F
\ee
for $1\le n\le7$, for which the flavor symmetry was derived in the detailed discussion for (\ref{t16}) (notice that the answer quoted there holds true for $n=7$ as well, but it is not displayed since it matches the classical flavor symmetry for $\su(6)_5+2\L^2$). For $n=7$, the geometry can be written as
\be

\ee
which manifests $\su(6)_5+\half\L^3+\L^2+\F$ description. Integrating out $z$ living in top-most compact surface leads to $\su(6)_\frac{11}2+\half\L^3+\L^2$.

\subsubsection*{\ubf{Derivation of (\ref{t65}) and (\ref{t66})}:}
These theories can be produced using the KK theory
\be
%
\ee
whose associated geometry can be written as
\be
%
\ee
The theories in (\ref{t65}) are produced by successively integrating out $f-z_i$ living in $\bF_0^{2+2+2+2}$. Integrating out $f-z_2$ removes $\mathbf{N}_0$ and $\mathbf{N}_3$. Subsequently integrating out $f-z_1$ removes $\mathbf{N}_1$ as well. This verifies the result claimed in (\ref{t65}).

The theories in (\ref{t66}) are produced by successively integrating out $z_i$ living in $\bF_0^{2+2+2+2}$. Integrating out $z_1$ removes $\mathbf{N}_0$ and $\mathbf{N}_4$. Subsequently integrating out $z_2$ removes $\mathbf{N}_1$ as well. This verifies the result claimed in (\ref{t66}).

\subsubsection*{\ubf{Derivation of (\ref{t67})}:}
These theories can be obtained from the KK theory
\be

\ee
Including the flavor symmetries, we can represent the $5d$ KK theory as
\be
%
\ee
Using the gluing rules, we can figure out the geometry to be
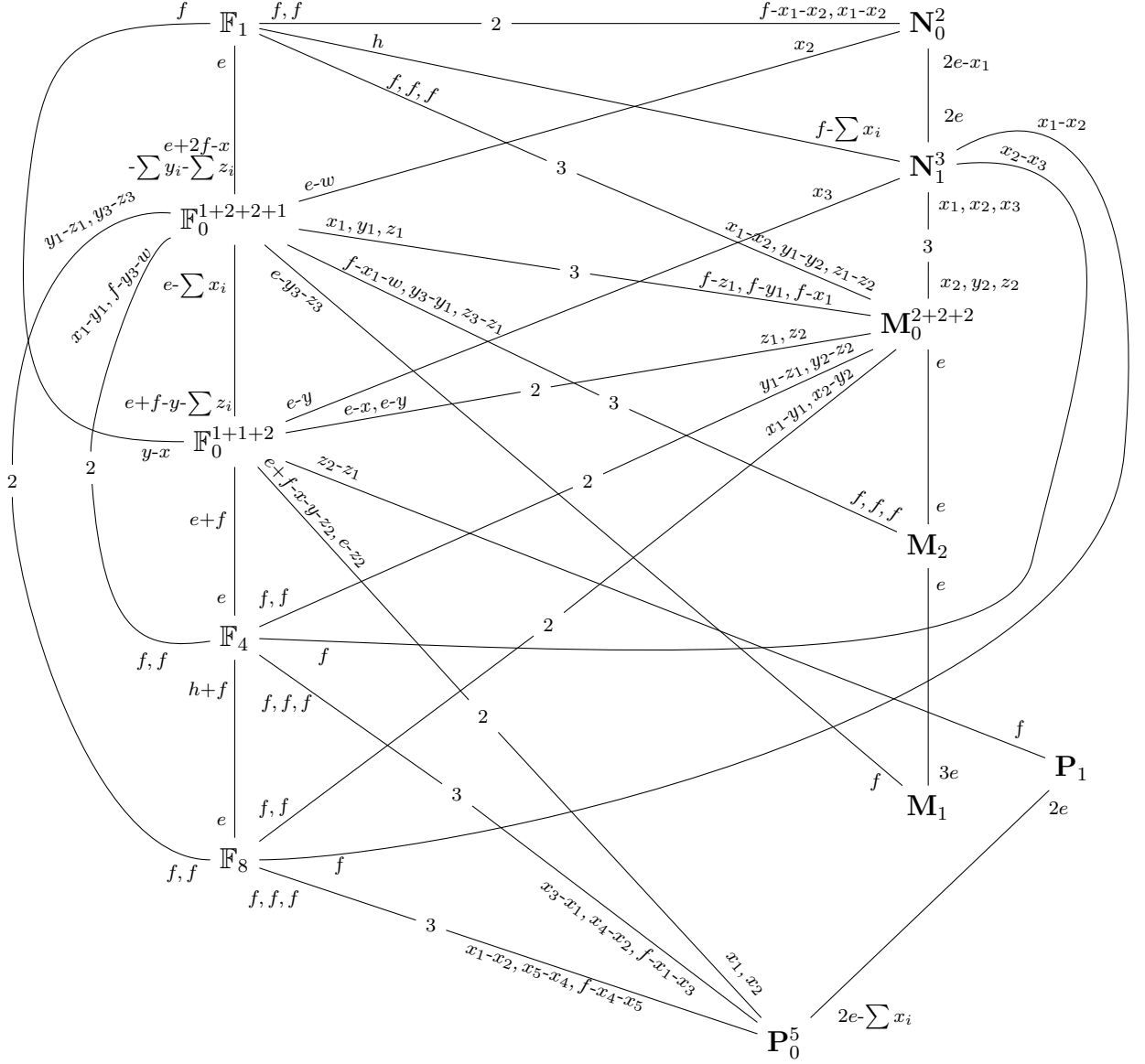
\be
%
\ee
The theories (\ref{t67}) are obtained by successively integrating out $f-z_i$ living in $\bF_0^{1+1+2}$. Integrating out $f-z_2$ removes $\mathbf{P}_1$, $\mathbf{N}_1$ and $\mathbf{M}_0$, leading to an $\su(2)^2\oplus\fg_2$ flavor symmetry. Further integrating out $f-z_1$ removes $\mathbf{P}_0$, thus leading to an $\su(2)\oplus\fg_2$ flavor symmetry. This verifies the results claimed in (\ref{t67}).

\subsubsection*{\ubf{Derivation of (\ref{t68}---\ref{t70})}:}
These can be produced using the KK theory
\be
%
\ee
for which the geometry is
\be
%
\ee
The theories in (\ref{t68}) are produced by integrating out $x_i$ living in $\bF_0^{10+2}$. The theories in (\ref{t69}) are produced by integrating out $x_i$ after integrating out $f-x_1$ living in $\bF_0^{10+2}$. The theories in (\ref{t70}) are produced by integrating out $x_i$ after integrating out $f-x_1,f-x_2$ living in $\bF_0^{10+2}$.

\subsubsection*{\ubf{Derivation of (\ref{t71})}:}
These theories can be derived from the KK theory
\be
%
\ee
The corresponding $6d$ SCFT has an $\fe_8$ flavor symmetry arising from the $\sp(0)$ node. 

The $\su(2)$ node carries 4 full hypers, out of which two half-hypers are trapped by the two $\su(1)$ nodes. As we know from before, if only one half-hyper is trapped, then the remaining 7 half-hypers transform as $\F$ of $\fg_2$. Now, since another half-hyper is trapped, we expect the remaining 6 half-hypers to transform as $\F\oplus\bar{\F}$ of $\su(3)$. This would suggest that the KK theory admits a collection of non-compact surfaces comprising $\su(3)^{(1)}$. However, this leads to a wrong prediction for the flavor symmetry for $\su(6)_1+\L^3+8\F$. That is, it predicts an $\so(16)\oplus\u(2)$ flavor symmetry, while we know from the analysis for (\ref{t68}) that the correct flavor symmetry is $\so(16)\oplus\su(2)^2$.

We claim that the $\su(3)$ flavor symmetry actually affinizes to $\fg_2^{(1)}$ rather than $\su(3)^{(1)}$, with the coupling of the corresponding non-compact surfaces shown below. This leads to the correct $\so(16)\oplus\so(4)$ flavor symmetry for $\su(6)_1+\L^3+8\F$ as we verify below. The geometry manifesting the KK theory description is the following
\be

\ee
The $\su(6)_\frac32+\L^3+9\F$ description is obtained by applying some flops and isomorphisms that can be found in \cite{Bhardwaj:2020gyu}. This description is manifested by the following geometry
\be
%
\ee
Now, the theory $\su(6)_1+\L^3+8\F$ can be obtained by integrating out $f-x_1$ living in $\bF_1^9$, which can be seen to lead to the removal of $\mathbf{N}_7$ and $\mathbf{M}_2$, implying that this theory has flavor symmetry $\so(16)\oplus\su(2)^2$. Thus, we see that if the affinization is $\fg_2^{(1)}$ instead of $\su(3)^{(1)}$, then we obtain the correct flavor symmetry.

The theories (\ref{t71}) can be obtained by successively integrating out $x_i$ living in $\bF_1^9$. The reader can check the results claimed in (\ref{t71}). For example, integrating out $x_9$ integrates out $\mathbf{N}_0$ and $\mathbf{M}_0$, indeed leaving an $\fe_8\oplus\fg_2$ flavor symmetry.

\subsubsection*{\ubf{Derivation of (\ref{t72}---\ref{t74})}:}
These theories can be produced using the KK theory
\be

\ee
The associated geometry can be written as
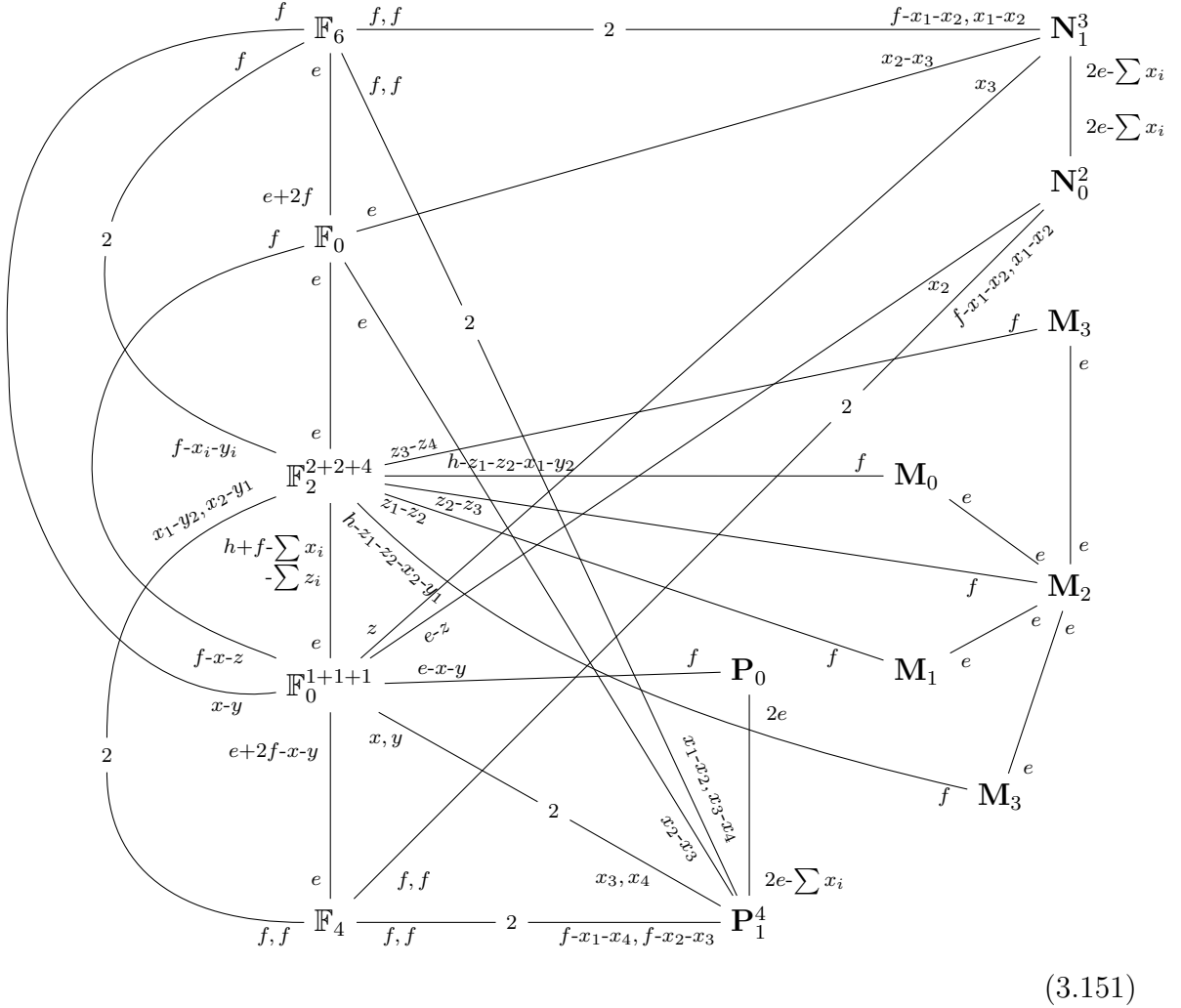
\be
%
\ee
The theories in (\ref{t72}) are obtained by integrating out $z_i$ living in $\bF_2^{2+2+4}$. The theories in (\ref{t73}) are obtained by integrating out $z_i$ after integrating out $f-z_1$ living in $\bF_2^{2+2+4}$. The theory in (\ref{t74}) is obtained by integrating out $f-z_1,f-z_2,z_4,z_3$ (in that order) living in the surface labeled as $\bF_2^{2+2+4}$ above.

\subsubsection*{\ubf{Derivation of (\ref{t75})}:}
These theories can be produced from the KK theory
\be
%
\ee
The $\sp(0)$ node in the corresponding $6d$ SCFT allows an $\su(3)\oplus\su(3)$ flavor symmetry. Twisting both gauge $\su(3)$ nodes by outer automorphism also twists both the flavor $\su(3)$ nodes by outer automorphism. The geometry is
\be
%
\ee
To obtain (\ref{t75}), we have to successively integrate out $x_i$ living in $\bF_6^{3+3}$.

\subsubsection*{\ubf{Derivation of (\ref{t76}) and (\ref{t77})}:}
These can be produced using the KK theory
\be
%
\ee
Drawing the geometry in a graphical form is quite tedious for this case. Thus, we only depict only partial data about the geometry graphically as follows
\be
%
\ee
where we have have labeled the compact surfaces as $\mathbf{i}_n^b$ which denotes $\bF_n^b$ and $\mathbf{i}$ is simply a label allowing us to refer to this surface as $\mathbf{S}_i$, which we shall do in what follows. We have also displayed all the $\P^1$ fibered non-compact surfaces. However, we have omitted all the ``mutual'' edges, that is edges between compact and non-compact surfaces, and edges between non-compact surfaces comprising different simple factors of the flavor symmetry algebra (or its affinized version). The data of these omitted edges is displayed in the following gluing rules:
\bit
\item $f-x_1,f-y_1$ in $\mathbf{S}_1$ are glued to $f-x,x$ in $\mathbf{N}_0$.
\item $x_i-x_{i+1},y_i-y_{i+1}$ in $\mathbf{S}_1$ are glued to $f,f$ in $\mathbf{N}_i$ for $i=1,\cdots,4$.
\item $x_5,y_5$ in $\mathbf{S}_1$ are glued to $x_1,x_1$ in $\mathbf{N}_5$.
\item $f$ in $\mathbf{S}_{i+1}$ is glued to $x_{i+1}-x_i$ in $\mathbf{N}_5$ for $i=1,2$.
\item $e$ in $\mathbf{S}_{4}$ is glued to $x_{4}-x_3$ in $\mathbf{N}_5$.
\item $f,f$ in $\mathbf{S}_5$ are glued to $f-x_4-x_5,x_5-x_4$ in $\mathbf{N}_5$.
\item $x_1,x_1,y_1,y_1,z_2,f-z_2,f-z_2-x_2,z_2-x_2,z_1,f-z_1,f-z_1-y_2,z_1-y_2$ in $\mathbf{S}_5$ are glued to $x_2,y_2,x_1,y_1,f-x_3,f-y_3,f,f,f-x_4,f-y_4,f,f$ in $\mathbf{M}_0$.
\item $f-y,f-y,f-x,f-x$ in $\mathbf{S}_3$ are glued to $x_3,y_3,x_4,y_4$ in $\mathbf{M}_0$.
\item $f-x,x,x,f-x$ in $\mathbf{S}_2$ are glued to $f-x_3,f-y_3,f-x_2,f-y_2$ in $\mathbf{M}_0$.
\item $f,f,f,f$ in $\mathbf{S}_1$ are glued to $x_2-x_1,y_2-y_1,x_3-x_4,y_3-y_4$ in $\mathbf{M}_0$.
\item $f-x_1,f-y_1,x_2,y_2$ in $\mathbf{S}_5$ are glued to $f-x_1,f-x_2,x_3,x_4$ in $\mathbf{M}_1$.
\item $e,e$ in $\mathbf{S}_4$ are glued to $x_1-x_3,x_2-x_4$ in $\mathbf{M}_1$.
\item $x,y$ in $\mathbf{S}_3$ are glued to $x_3,x_4$ in $\mathbf{M}_1$.
\item $f$ in $\mathbf{S}_2$ is glued to $f-x_1-x_3$ in $\mathbf{M}_1$.
\item $f,f$ in $\mathbf{S}_1$ are glued to $x_1-x_2,x_3-x_4$ in $\mathbf{M}_1$.
\eit
Note that the gluings above should be read in the order they are presented. For instance, ``$C_1,C_2$ is glued to $D_1,D_2$'' means that $C_1$ is glued to $D_1$ and $C_2$ is glued to $D_2$.

The theories in (\ref{t76}) are produced by successively integrating out $x_i$ living in $\mathbf{S}_1$. It is easy to see that during these processes $\mathbf{M}_1$ is integrated out but $\mathbf{M}_0$ always survives. Thus there is always an $\su(2)$ factor present in the flavor symmetry. The other non-abelian factors arise from the collection of surfaces $\mathbf{N}_i$ and can be easily figured out from the above gluing rules.

To construct the theories in (\ref{t77}), we need to perform a few flops which lead to the following representation of the above geometry
\be

\ee
along with the following gluing rules:
\bit
\item $f,f$ in $\mathbf{S}_1$ are glued to $f-x_1-x_2,x_1-x_2$ in $\mathbf{N}_0$.
\item $e-x_5$ in $\mathbf{S}_2$ is glued to $x_2$ in $\mathbf{N}_0$.
\item $x_{6-i}-x_{5-i}$ in $\mathbf{S}_2$ is glued to $f$ in $\mathbf{N}_i$ for $i=1,\cdots,4$.
\item $x_1$ in $\mathbf{S}_2$ is glued to $x_1$ in $\mathbf{N}_5$.
\item $f$ in $\mathbf{S}_{3}$ is glued to $x_{2}-x_1$ in $\mathbf{N}_5$.
\item $e$ in $\mathbf{S}_{4}$ is glued to $x_{3}-x_2$ in $\mathbf{N}_5$.
\item $f,f$ in $\mathbf{S}_5$ are glued to $f-x_3-x_4,x_4-x_3$ in $\mathbf{N}_5$.
\item $z_2,f-z_2,f-z_2-x,z_2-x,z_1,f-z_1,f-z_1-y,z_1-y$ in $\mathbf{S}_5$ are glued to $f-x_1,f-y_1,f,f,f-x_2,f-y_2,f,f$ in $\mathbf{M}_0$.
\item $f-y,f-y,f-x,f-x$ in $\mathbf{S}_3$ are glued to $x_1,y_1,x_2,y_2$ in $\mathbf{M}_0$.
\item $y_1,f-y_1,y_2-y_1,f-y_2-y_1$ in $\mathbf{S}_2$ are glued to $f-x_1,f-y_1,f,f$ in $\mathbf{M}_0$.
\item $f,f$ in $\mathbf{S}_1$ are glued to $x_1-x_2,y_1-y_2$ in $\mathbf{M}_0$.
\item $f,f,x,y$ in $\mathbf{S}_5$ are glued to $f-x_1-x_6,f-x_2-x_5,x_3,x_4$ in $\mathbf{M}_1$.
\item $e,e$ in $\mathbf{S}_4$ are glued to $x_1-x_3,x_2-x_4$ in $\mathbf{M}_1$.
\item $x,y$ in $\mathbf{S}_3$ are glued to $x_3,x_4$ in $\mathbf{M}_1$.
\item $e-y_2,e$ in $\mathbf{S}_2$ are glued to $x_6,f-x_1-x_3$ in $\mathbf{M}_1$.
\item $f,f,f$ in $\mathbf{S}_1$ are glued to $x_1-x_2,x_3-x_4,x_5-x_6$ in $\mathbf{M}_1$.
\eit
The theories in (\ref{t77}) are obtained by successively integrating out $x_i$ after integrating out $f-x_5$ living in $\mathbf{S}_2$. The reader can easily verify that this leads to the results claimed in (\ref{t77}). For example, first integrating out $f-x_5$ integrates out $\mathbf{N}_1$ and $\mathbf{M}_1$, and then integrating out $x_1$ integrates out $\mathbf{N}_4$, thus leading to the conclusion that the $n=1$ case of (\ref{t77}) has $\su(2)^3\oplus\su(3)$ flavor symmetry.

\subsubsection*{\ubf{Derivation of (\ref{t78})}:}
This theory can be obtained by using the KK theory
\be

\ee
The geometry can be written as
\be
%
\ee
along with the following gluing rules:
\bit
\item $f,f,f,f,f,f$ in $\mathbf{S}_1$ are glued to $x_4-x_5,y_4-y_5,x_6-x_7,x_8-x_9,y_6-y_7,y_8-y_9$ in $\mathbf{N}_0$.
\item $x_3,f-x_3,x_2,x_2,x_1,x_1,x_4-x_3,f-x_4-x_3$ in $\mathbf{S}_2$ are glued to $f-x_4,f-y_4,x_7,y_7,x_9,y_9,f,f$ in $\mathbf{N}_0$.
\item $f,f,f,f$ in $\mathbf{S}_3$ are glued to $x_5-x_7,x_4-x_6,y_5-y_7,y_4-y_6$ in $\mathbf{N}_0$.
\item $f,f,f,f$ in $\mathbf{S}_4$ are glued to $x_7-x_9,x_6-x_8,y_7-y_9,y_6-y_8$ in $\mathbf{N}_0$.
\item $f-x_1,x_1,f-x_2,x_2,x_1,f-x_1,x_2,f-x_2$ in $\mathbf{S}_{5}$ is glued to $f-x_6,f-x_5,f-x_7,f-x_4,f-y_6,f-y_5,f-y_7,f-y_4$ in $\mathbf{N}_0$.
\item $f,f,f,f$ in $\mathbf{S}_1$ are glued to $x_1-x_4,x_2-x_5,x_3-x_6,x_7-x_8$ in $\mathbf{N}_1$.
\item $e-x_1,e-x_2,e-x_4,e-x_5$ in $\mathbf{S}_2$ are glued to $f-x_3,f-x_2,f-x_1,x_8$ in $\mathbf{N}_1$.
\item $f,f$ in $\mathbf{S}_3$ are glued to $x_3-x_7,x_6-x_8$ in $\mathbf{N}_1$.
\item $f,f$ in $\mathbf{S}_4$ are glued to $x_5-x_6,x_2-x_3$ in $\mathbf{N}_1$.
\item $f,f$ in $\mathbf{S}_5$ are glued to $x_4-x_5,x_1-x_2$ in $\mathbf{N}_1$.
\item $f,f,f,f,f,f,f,f$ in $\mathbf{S}_1$ are glued to $y_1-y_4,y_2-y_5,y_3-y_6,y_7-y_8,z_1-z_4,z_2-z_5,z_3-z_6,z_7-z_8$ in $\mathbf{N}_2$.
\item $e+f-x_5,e+f-x_5,e+f-x_5,e-x_5,e-x_5$ in $\mathbf{S}_2$ are glued to $x_5-y_1-y_2,x_6-z_1-y_3,x_8-z_2-z_3,y_8,z_8$ in $\mathbf{N}_2$.
\item $f,f,f,f,f,f,f,f$ in $\mathbf{S}_3$ are glued to $f-x_1-x_5,x_1-x_5,x_4-x_6,x_7-x_8,y_6-y_8,y_3-y_7,z_6-z_8,z_3-z_7$ in $\mathbf{N}_2$.
\item $f,f,f,f,f,f,f,f$ in $\mathbf{S}_4$ are glued to $f-x_1-x_4,x_1-x_4,x_3-x_7,x_5-x_6,y_5-y_6,y_2-y_3,z_5-z_6,z_2-z_3$ in $\mathbf{N}_2$.
\item $f,f,f,f,f,f,f,f$ in $\mathbf{S}_5$ are glued to $f-x_2-x_3,x_2-x_3,x_4-x_7,x_6-x_8,y_4-y_5,y_1-y_2,z_4-z_5,z_1-z_2$ in $\mathbf{N}_2$.
\eit
The theory (\ref{t78}) is produced by integrating out $f-x_5$ from $\mathbf{S}_2$ which integrates out $\mathbf{N}_1$, thus leaving an $\su(2)^2$ flavor symmetry.

\subsubsection*{\ubf{Derivation of (\ref{t79})}:}
These can be constructed using the KK theory
\be

\ee
for which the geometry is
\be
%
\ee
The theories in (\ref{t79}) are obtained by successively integrating out $x_i$ living in $\bF_6^{7+7}$.

\subsubsection*{\ubf{Derivation of (\ref{t80})}:}
This theory can be obtained from the KK theory
\be
%
\ee
for which the geometry can be figured out to be
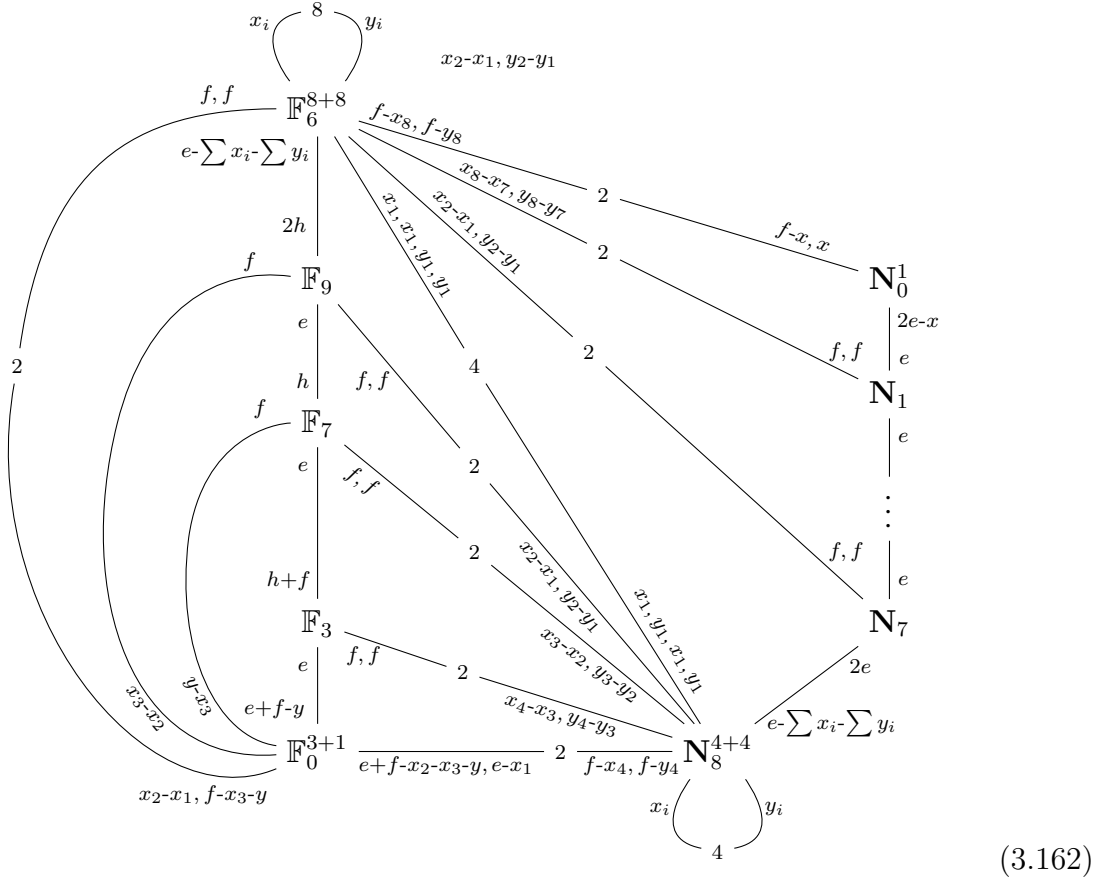
\be
%
\ee
The fundamentals can be integrated out by successively integrating out $x_i$ living in $\bF_6^{8+8}$. Integrating out $x_1$ integrates out $\mathbf{N}_1$ thus leaving an $\su(2)\oplus\sp(7)$ flavor symmetry as claimed in (\ref{t80}). Further integrating out $x_2$, integrates out $\mathbf{N}_6$ and $\mathbf{N}_8$ thus leading to $\sp(6)\oplus\u(1)$ flavor symmetry, which shows no enhancement. Thus, there is no enhancement as we integrate out even more fundamentals.

\subsubsection*{\ubf{Derivation of (\ref{t81})}:}
These theories can be obtained from the KK theory
\be

\ee
for which the geometry can be written as
\be
%
\ee
The theories in (\ref{t81}) are produced by successively integrating out $y_i$ living in $\bF_0^{2+2}$.

\subsubsection*{\ubf{Derivation of (\ref{t82})}:}
This theory can be produced using the KK theory
\be
%
\ee
The geometry for this theory can be figured out to be
\be
%
\ee
along with the following gluing rules:
\bit
\item $f,f$ in $\mathbf{S}_1$ are glued to $f-x_1-x_2,x_1-x_2$ in $\mathbf{N}_4$.
\item $e-z_4$ in $\mathbf{S}_{2}$ is glued to $x_2$ in $\mathbf{N}_4$.
\item $z_{i+1}-z_i$ in $\mathbf{S}_2$ is glued to $f$ in $\mathbf{N}_i$ for $i=1,2,3$.
\item $z_1$ in $\mathbf{S}_{2}$ is glued to $x_3$ in $\mathbf{N}_0$.
\item $f$ in $\mathbf{S}_{3}$ is glued to $x_2-x_3$ in $\mathbf{N}_0$.
\item $f$ in $\mathbf{S}_{4}$ is glued to $f-x_1-x_2$ in $\mathbf{N}_0$.
\item $f$ in $\mathbf{S}_{5}$ is glued to $x_1-x_2$ in $\mathbf{N}_0$.
\item $f,f,f,f$ in $\mathbf{S}_1$ are glued to $x_1-x_2,x_3-x_5,x_4-x_6,x_7-x_8$ in $\mathbf{M}_3$.
\item $e,x_3,y_3$ in $\mathbf{S}_2$ are glued to $x_2-x_3,x_6,x_8$ in $\mathbf{M}_3$.
\item $f,f$ in $\mathbf{S}_3$ are glued to $x_3-x_4,x_5-x_6$ in $\mathbf{M}_3$.
\item $f,f$ in $\mathbf{S}_4$ are glued to $f-x_1-x_5,f-x_2-x_3$ in $\mathbf{M}_3$.
\item $f,f$ in $\mathbf{S}_5$ are glued to $x_4-x_7,x_6-x_8$ in $\mathbf{M}_3$.
\item $x_2-x_3,y_2-y_3$ in $\mathbf{S}_2$ are glued to $f,f$ in $\mathbf{M}_2$.
\item $x_1-x_2,y_1-y_2$ in $\mathbf{S}_2$ are glued to $f,f$ in $\mathbf{M}_1$.
\item $e-x_1-y_1$ in $\mathbf{S}_{2}$ is glued to $f$ in $\mathbf{M}_0$.
\eit
The fundamentals are integrated out by successively integrating out $f-z_i$ living in $\mathbf{S}_2$. Integrating out $f-z_4$ integrates out $\mathbf{N}_3$, $\mathbf{M}_0$ and $\mathbf{M}_3$, thus leaving a $\u(3)\oplus\sp(3)\oplus\su(2)$ flavor symmetry as claimed in (\ref{t82}). Now, further integrating out $f-z_3$ further integrates out $\mathbf{N}_2$ and $\mathbf{N}_4$, thus leaving a $\u(3)\oplus\sp(2)\oplus\u(1)$ flavor symmetry, which is just the classical flavor symmetry. Hence, integrating more than one fundamental out of the KK theory leaves only a classical flavor symmetry without any enhancement.

\subsubsection*{\ubf{Derivation of (\ref{t83})}:}
These theories can be produced using the KK theory
\be

\ee
for which the geometry can be figured out to be
\be
%
\ee
The fundamentals are integrated out by successively integrating out $x_i$ living in $\bF_8^{6+6}$.

\subsubsection*{\ubf{Derivation of (\ref{t84})}:}
This theory can be constructed using the KK theory
\be
%
\ee
The geometry can be figured out to be
\be
%
\ee
along with the following gluing rules:
\bit
\item $e+f-\sum x_i-y$ in $\mathbf{S}_1$ is glued to $x_4$ in $\mathbf{N}_0$.
\item $f$ in $\mathbf{S}_{i+1}$ is glued to $x_{4-i}-x_{5-i}$ in $\mathbf{N}_0$ for $i=1,2,3$.
\item $x_1,f-y_1$ in $\mathbf{S}_5$ are glued to $f-x_1,f-x_2$ in $\mathbf{N}_0$.
\item $e-z$ in $\mathbf{S}_1$ is glued to $x_4$ in $\mathbf{N}_1$.
\item $f$ in $\mathbf{S}_{i+1}$ is glued to $x_{4-i}-x_{5-i}$ in $\mathbf{N}_1$ for $i=1,2,3$.
\item $x_1,f-y_1$ in $\mathbf{S}_5$ are glued to $f-x_1,f-x_2$ in $\mathbf{N}_1$.
\item $x_{i}-x_{i-1},y_{i-1}-y_{i}$ in $\mathbf{S}_5$ are glued to $f,f$ in $\mathbf{N}_i$ for $i=2,\cdots,7$.
\item $y_7-x_7$ in $\mathbf{S}_5$ is glued to $f$ in $\mathbf{N}_8$.
\eit
The fundamentals are integrated out by successively integrating out $x_i$ living in $\mathbf{S}_5$.

\subsection{Rank 6}
\subsubsection*{\ubf{Derivation of (\ref{t85}---\ref{t87})}:}
These can be produced using the KK theory
\be

\ee
for which the geometry can be figured out to be
\be
%
\ee
along with the following gluing rules:
\bit
\item $e$ in $\mathbf{S}_4$ is glued to $x-y$ in $\mathbf{N}_6$.
\item $y_5,y_6$ in $\mathbf{S}_{3}$ are glued to $f-x,y$ in $\mathbf{N}_6$.
\item $y_i-y_{i+1}$ in $\mathbf{S}_3$ is glued to $f$ in $\mathbf{N}_i$ for $i=1,\cdots,5$.
\item $e-y_1-y_2$ in $\mathbf{S}_3$ is glued to $f$ in $\mathbf{N}_0$.
\item $f,f$ in $\mathbf{S}_{1}$ are glued to $f-x_1-x_2,x_1-x_2$ in $\mathbf{M}_0$.
\item $e-x$ in $\mathbf{S}_2$ is glued to $x_2$ in $\mathbf{M}_0$.
\item $x$ in $\mathbf{S}_2$ is glued to $x_3$ in $\mathbf{M}_1$.
\item $e$ in $\mathbf{S}_5$ is glued to $x_2-x_3$ in $\mathbf{M}_1$.
\item $f,f$ in $\mathbf{S}_6$ are glued to $x_1-x_2,f-x_1-x_2$ in $\mathbf{M}_1$.
\eit
The theories in (\ref{t85}) are produced by integrating out $y_i$ living in $\mathbf{S}_3$. The theories in (\ref{t86}) are produced by integrating out $y_i$ after integrating out $f-y_1$ living in $\mathbf{S}_3$. The theories in (\ref{t87}) are produced by integrating out (at least two) $y_i$ after integrating out $f-y_1,f-y_2$ living in $\mathbf{S}_3$.

\subsubsection*{\ubf{Derivation of (\ref{t88})}:}
These can be produced using the KK theory
\be

\ee
for which the geometry can be figured out to be
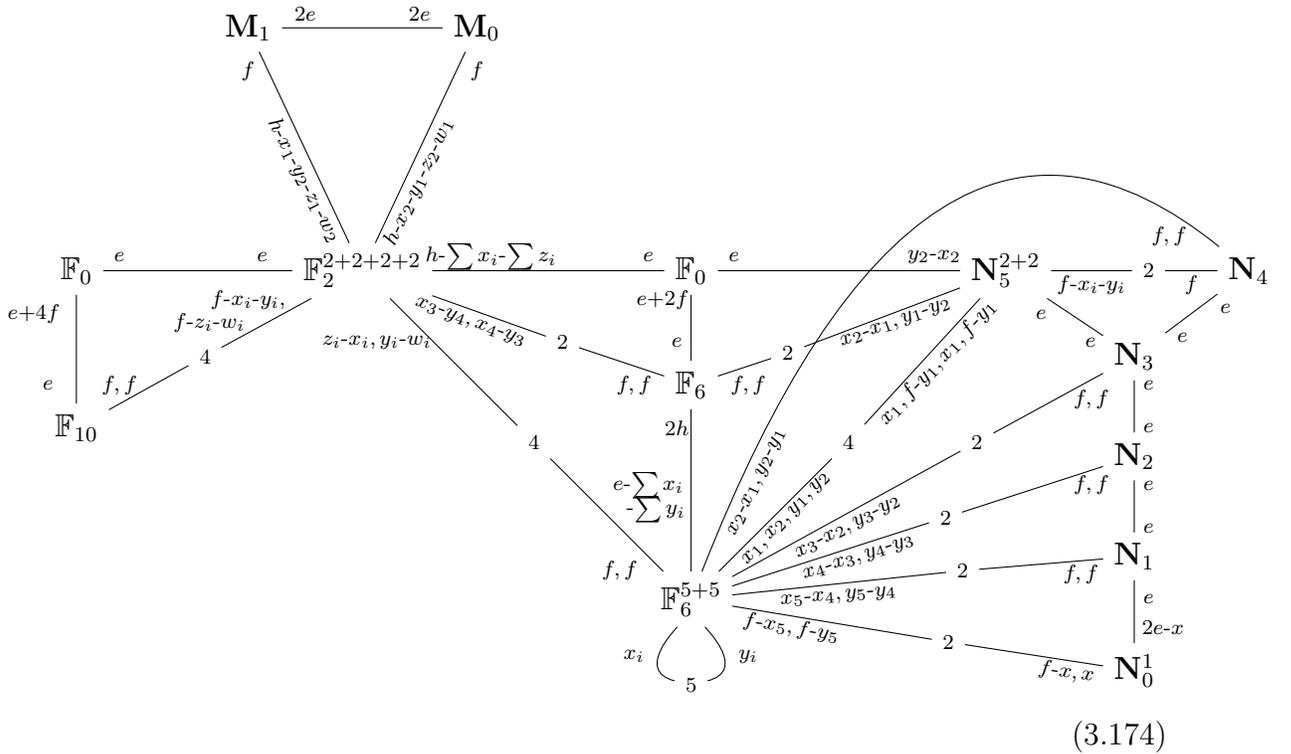
\be
%
\ee
The theories in (\ref{t88}) are produced by integrating out $x_i$ in $\bF_6^{5+5}$.

\subsubsection*{\ubf{Derivation of (\ref{t89})}:}
These can be produced using the KK theory
\be
%
\ee
The associated geometry can be written as
\be
%
\ee
along with the following gluing rules:
\bit
\item $f-x_9,f-y_9$ in $\mathbf{S}_1$ are glued to $f-x,x$ in $\mathbf{N}_9$.
\item $x_{i+1}-x_{i},y_{i+1}-y_{i}$ in $\mathbf{S}_1$ are glued to $f,f$ in $\mathbf{N}_i$ for $i=1,\cdots,8$.
\item $x_1,x_2,y_1,y_2$ in $\mathbf{S}_1$ are glued to $x_1,f-y_1,x_1,f-y_1$ in $\mathbf{N}_0$.
\item $f,f$ in $\mathbf{S}_i$ are glued to $x_{i}-x_{i-1},y_{i-1}-y_{i}$ in $\mathbf{N}_0$ for $i=2,3,4$.
\item $e$ in $\mathbf{S}_5$ is glued to $y_4-x_4$ in $\mathbf{N}_0$.
\eit
The theories in (\ref{t89}) are produced by successively integrating out $x_i$ living in $\mathbf{S}_1$.

\subsubsection*{\ubf{Derivation of (\ref{t90})}:}
These can be produced using the KK theory
\be

\ee
whose geometry can be figured out to be
\be
%
\ee
The theories in (\ref{t90}) can be produced by successively integrating out $y_i$ living in the right-most compact surface.

\subsubsection*{\ubf{Derivation of (\ref{t91})}:}
This theory can be produced using the KK theory
\be
%
\ee
for which the geometry can be written as
\be
%
\ee
along with the following gluing rules:
\bit
\item $h-x_1-y_2-z_2-w_1$ in $\mathbf{S}_6$ is glued to $f$ in $\mathbf{N}_0$.
\item $h-x_2-y_1-z_1-w_1$ in $\mathbf{S}_6$ is glued to $f$ in $\mathbf{N}_1$.
\item $w_1-w_2$ in $\mathbf{S}_6$ is glued to $f$ in $\mathbf{N}_2$.
\item $f-x_1,x_1,x_1,f-x_1,f-x_2,x_2,x_2,f-x_2$ in $\mathbf{S}_1$ are glued to $f-x_3,f-y_3,f-x_2,f-y_2,f-x_4,f-y_4,f-x_1,f-y_1$ in $\mathbf{M}_0$.
\item $f,f,f,f$ in $\mathbf{S}_2$ are glued to $x_4-x_6,y_4-y_6,x_3-x_5,y_3-y_5$ in $\mathbf{M}_0$.
\item $f,f,f,f$ in $\mathbf{S}_3$ are glued to $x_2-x_4,y_2-y_4,x_1-x_3,y_1-y_3$ in $\mathbf{M}_0$.
\item $x_3,f-x_3,x_2,x_2,x_1,x_1,x_4-x_3,f-x_4-x_3$ in $\mathbf{S}_4$ are glued to $f-x_1,f-y_1,x_4,y_4,x_6,y_6,f,f$ in $\mathbf{M}_0$.
\item $f,f,f,f,f,f$ in $\mathbf{S}_5$ are glued to $x_1-x_2,y_1-y_2,x_3-x_4,y_3-y_4,x_5-x_6,y_5-y_6$ in $\mathbf{M}_0$.
\item $f,f,f,f$ in $\mathbf{S}_1$ are glued to $f-x_2-x_3,x_2-x_3,x_4-x_7,x_6-x_8$ in $\mathbf{M}_1$.
\item $f,f,f,f$ in $\mathbf{S}_2$ are glued to $f-x_1-x_4,x_1-x_4,x_3-x_7,x_5-x_6$ in $\mathbf{M}_1$.
\item $f,f,f,f$ in $\mathbf{S}_3$ are glued to $f-x_1-x_5,x_1-x_5,x_4-x_6,x_7-x_8$ in $\mathbf{M}_1$.
\item $e-x_1,e-x_2,e-x_4$ in $\mathbf{S}_4$ are glued to $x_5,x_6,x_8$ in $\mathbf{M}_1$.
\eit
The theory in (\ref{t91}) is produced by integrating out $w_2$ living in $\mathbf{S}_6$ which integrates out $\mathbf{N}_2$ and $\mathbf{M}_1$, thus leaving an $\su(3)\oplus\su(2)$ flavor symmetry as claimed in (\ref{t91}).

\subsubsection*{\ubf{Derivation of (\ref{t92})}:}
These can be produced using the KK theory
\be

\ee
for which the geometry can be written as
\be
%
\ee
along with the following gluing rules:
\bit
\item $h-x_1-y_2-z_2-w_2$ in $\mathbf{S}_1$ is glued to $f$ in $\mathbf{M}_0$.
\item $h-x_2-y_1-z_1-w_1$ in $\mathbf{S}_1$ is glued to $f$ in $\mathbf{M}_1$.
\item $e$ in $\mathbf{S}_2$ is glued to $x_4-y_4$ in $\mathbf{N}_0$.
\item $f,f$ in $\mathbf{S}_{2+i}$ are glued to $x_{4-i}-x_{5-i},y_{5-i}-y_{4-i}$ in $\mathbf{N}_0$ for $i=1,2,3$.
\item $y_2,y_1,f-x_2,f-x_1$ in $\mathbf{S}_6$ are glued to $f-x_1,y_1,f-x_2,y_2$ in $\mathbf{N}_0$.
\item $x_{i}-x_{i+1},y_{i+1}-y_{i}$ in $\mathbf{S}_{6}$ are glued to $f,f$ in $\mathbf{N}_i$ for $i=1,\cdots,7$.
\item $x_8-y_8$ in $\mathbf{S}_6$ is glued to $f$ in $\mathbf{N}_8$.
\eit
The theories in (\ref{t92}) are produced by integrating out $y_i$ living in $\mathbf{S}_6$.

\subsubsection*{\ubf{Derivation of (\ref{t93})}:}
These can be produced using the KK theory
\be

\ee
for which the geometry can be written as
\be
%
\ee
along with the following gluing rules:
\bit
\item $2e+f-\sum x_i-y$ in $\mathbf{S}_1$ is glued to $x_5-y_5$ in $\mathbf{N}_0$.
\item $f,f$ in $\mathbf{S}_{1+i}$ are glued to $x_{5-i}-x_{6-i},y_{6-i}-y_{5-i}$ in $\mathbf{N}_0$ for $i=1,\cdot,4$.
\item $y_2,y_1,f-x_2,f-x_1$ in $\mathbf{S}_6$ are glued to $f-x_1,y_1,f-x_2,y_2$ in $\mathbf{N}_0$.
\item $x_{i}-x_{i+1},y_{i+1}-y_{i}$ in $\mathbf{S}_{6}$ are glued to $f,f$ in $\mathbf{N}_i$ for $i=1,\cdots,8$.
\item $x_9-y_9$ in $\mathbf{S}_6$ is glued to $f$ in $\mathbf{N}_9$.
\eit
The theories in (\ref{t93}) are produced by integrating out $y_i$ living in $\mathbf{S}_6$.

\subsubsection*{\ubf{Derivation of (\ref{t94})}:}
This theory can be produced using the KK theory
\be

\ee
for which the geometry can be written as
\be
%
\ee
The $\F$ are integrated out by successively integrating out $f-x_i$ living in $\bF_0^4$. Integrating out $f-x_4$ integrates out $\mathbf{N}_3$ and $\mathbf{M}_1$ leaving an $\sp(4)\oplus\su(2)^2$ flavor symmetry as claimed in (\ref{t94}).

\subsubsection*{\ubf{Derivation of (\ref{t95}) and (\ref{t96})}:}
These theories can be produced using the KK theory
\be
%
\ee
The corresponding $6d$ SCFT on its tensor branch reduces to $\so(11)+6\F+\frac32\S$, thus implying an $\sp(6)\oplus\su(2)$ flavor symmetry for the $6d$ SCFT. The geometry for the KK theory can be written as
\be\label{G1}
%
\ee
where we have manifested the $\so(12)+\frac32\S+6\F$ frame and have omitted the non-compact surfaces corresponding to $\su(2)^{(1)}$. The gluing rules between $\mathbf{S}_i$ and $\mathbf{N}_j$ are:
\bit
\item $f,f$ in $\mathbf{S}_1$ are glued to $f-x_1-x_2,x_1-x_2$ in $\mathbf{N}_6$.
\item $e-x_6$ in $\mathbf{S}_2$ is glued to $x_2$ in $\mathbf{N}_6$.
\item $x_{i+1}-x_{i}$ in $\mathbf{S}_2$ is glued to $f$ in $\mathbf{N}_i$ for $i=1,\cdots,5$.
\item $x_1$ in $\mathbf{S}_2$ is glued to $x_4$ in $\mathbf{N}_0$.
\item $f$ in $\mathbf{S}_{2+i}$ is glued to $x_{4-i}-x_{5-i}$ in $\mathbf{N}_0$ for $i=1,2,3$.
\item $f$ in $\mathbf{S}_{6}$ is glued to $f-x_1-x_2$ in $\mathbf{N}_0$.
\eit
It was shown in \cite{Bhardwaj:2020gyu} that the above geometry is flop equivalent to the following geometry
\be\label{G2}

\ee
which manifests the $\so(12)+\S+\half\C+6\F$ frame and we have again omitted the non-compact surfaces corresponding to $\su(2)^{(1)}$. The gluing rules between $\mathbf{S}_i$ and $\mathbf{N}_j$ are the same as above.

Now, the gluing of flavor $\su(2)^{(1)}$ to a geometry for the KK theory
\be\label{e8}
%
\ee
was presented in Part 1. The geometry presented there can be turned into the above geometry (\ref{G2}) by first performing some perturbative flops and finally applying $\cS$ upon the surface labeled as $\mathbf{S}_2$ in (\ref{G2}). In this way, the coupling of flavor $\su(2)^{(1)}$ to the compact surfaces $\mathbf{S}_i$ in (\ref{G2}) can be figured out.

The $\F$ are integrated out from $\so(12)+\S+\half\C+6\F$ if we successively integrate out $f-x_i$ living in $\mathbf{S}_2$ of (\ref{G2}). We claim that the coupling of flavor $\su(2)^{(1)}$ is such that both the non-compact $\P^1$ fibered surfaces comprising $\su(2)^{(1)}$ are integrated out. Thus, the non-abelian contribution to the flavor symmetry for $5d$ SCFTs $\so(12)+\S+\half\C+(6-n)\F$ comes purely from the surfaces $\mathbf{N}_i$ in (\ref{G2}). This leads to the result presented in (\ref{t96}).

Now, to obtain the coupling of flavor $\su(2)^{(1)}$ to compact surfaces in (\ref{G1}) (starting from the coupling of flavor $\su(2)^{(1)}$ to the KK theory (\ref{e8}) presented in Part 1) requires performing a lot of non-trivial, complicated flops. Fortunately, the knowledge of precise coupling is not required to deduce the flavor symmetry for $5d$ SCFTs $\so(12)+\frac32\S+(6-n)\F$. For this deduction, first note that the non-abelian part of the flavor symmetry of a $5d$ SCFT must be given by a finite semi-simple Lie algebra. Thus, as we integrate out the first $\F$, at least one of the two non-compact surfaces comprising $\su(2)^{(1)}$ must be integrated out. 

The first $\F$ is integrated out by integrating out $f-x_6$ living in $\mathbf{S}_2$ of (\ref{G1}) which integrates out $\mathbf{N}_5$ leading to an $\su(2)\oplus\sp(5)$ contribution to the non-abelian part of the flavor symmetry of $5d$ SCFT $\so(12)+\frac32+5\F$. If this process integrates out only one of the non-compact surfaces comprising $\su(2)^{(1)}$, then the full flavor symmetry for $\so(12)+\frac32+5\F$ would be $\sp(5)\oplus\su(2)^2$ since an extra $\su(2)$ would be contributed to the non-abelian part of the flavor symmetry. If, on the other hand, this process integrates out both of the non-compact surfaces comprising $\su(2)^{(1)}$, then the full flavor symmetry for $\so(12)+\frac32+5\F$ would be $\sp(5)\oplus\u(2)$ since no other factor would be contributed to the non-abelian part of the flavor symmetry.

We claim that only one of the $\su(2)^{(1)}$ surfaces is integrated out and that (\ref{t95}) shows the correct flavor symmetry for this theory. To show this, let us assume, to the contrary, that the flavor symmetry for $\so(12)+\frac32+5\F$ is $\sp(5)\oplus\u(2)$, that is the only non-compact $\P^1$ fibered surfaces arising in the geometry for $\so(12)+\frac32+5\F$ are $\mathbf{N}_i$ for $i=0,\cdots,4$ and $\mathbf{N}_6$. Let us integrate out another fundamental to obtain $\so(12)+\frac32+4\F$. This is done by integrating out $f-x_6,f-x_5$ (in that order) living in $\mathbf{S}_2$ of (\ref{G1}). We see that this process only leaves non-compact surfaces $\mathbf{N}_i$ for $i=0,\cdots,3$ intact thus implying that the non-abelian part of the flavor symmetry for $\so(12)+\frac32+4\F$ is $\sp(4)$, but this is a contradiction since the non-abelian part of the classical flavor symmetry for $\so(12)+\frac32+4\F$ is $\sp(4)\oplus\su(2)$.

\subsubsection*{\ubf{Derivation of (\ref{t97})}:}
These can be produced using the KK theory
\be

\ee
for which the geometry can be written as
\be
%
\ee
The theories in (\ref{t97}) are produced by successively integrating out $f-x_i$ living in $\bF_0^4$.

\subsubsection*{\ubf{Derivation of (\ref{t98})}:}
These can be produced using the KK theory
\be
%
\ee
for which the geometry can be written as
\be
%
\ee
along with the following gluing rules:
\bit
\item $h-x_1-y_2-z_2-w_2$ in $\mathbf{S}_1$ is glued to $f$ in $\mathbf{M}_0$.
\item $h-x_2-y_1-z_1-w_1$ in $\mathbf{S}_1$ is glued to $f$ in $\mathbf{M}_1$.
\item $e$ in $\mathbf{S}_2$ is glued to $x_4-y_4$ in $\mathbf{N}_0$.
\item $f,f$ in $\mathbf{S}_{2+i}$ are glued to $x_{4-i}-x_{5-i},y_{5-i}-y_{4-i}$ in $\mathbf{N}_0$ for $i=1,2,3$.
\item $y_2,y_1,f-x_2,f-x_1$ in $\mathbf{S}_6$ are glued to $f-x_1,y_1,f-x_2,y_2$ in $\mathbf{N}_0$.
\item $x_{i}-x_{i+1},y_{i+1}-y_{i}$ in $\mathbf{S}_{6}$ are glued to $f,f$ in $\mathbf{N}_i$ for $i=1,\cdots,7$.
\item $x_8-y_8$ in $\mathbf{S}_6$ is glued to $f$ in $\mathbf{N}_8$.
\eit
The theories in (\ref{t98}) are produced by integrating out $y_i$ living in $\mathbf{S}_6$.

\subsection{Rank 7}
\subsubsection*{\ubf{Derivation of (\ref{t99})}:}
These can be produced using the KK theory
\be

\ee
for which the geometry can be figured out to be
\be
%
\ee
along with the following gluing rules:
\bit
\item $h-x_1-y_2-z_1-w_2$ in $\mathbf{S}_3$ is glued to $f$ in $\mathbf{M}_0$.
\item $h-x_2-y_1-z_2-w_1$ in $\mathbf{S}_3$ is glued to $f$ in $\mathbf{M}_1$.
\item $e$ in $\mathbf{S}_4$ is glued to $x_3-y_3$ in $\mathbf{N}_0$.
\item $f,f$ in $\mathbf{S}_{4+i}$ are glued to $x_{3-i}-x_{4-i},y_{4-i}-y_{3-i}$ in $\mathbf{N}_0$ for $i=1,2$.
\item $y_2,y_1,f-x_2,f-x_1$ in $\mathbf{S}_7$ are glued to $f-x_1,y_1,f-x_2,y_2$ in $\mathbf{N}_0$.
\item $x_{i}-x_{i+1},y_{i+1}-y_{i}$ in $\mathbf{S}_{7}$ are glued to $f,f$ in $\mathbf{N}_i$ for $i=1,\cdots,5$.
\item $x_6-y_6$ in $\mathbf{S}_7$ is glued to $f$ in $\mathbf{N}_6$.
\eit
The theories in (\ref{t99}) are produced by integrating out $y_i$ living in $\mathbf{S}_7$.

\section*{Acknowledgements}
The author thanks Sakura Schäfer-Nameki and Gabi Zafrir for discussions. The author is grateful to Julius Eckhard and Sakura Schäfer-Nameki for providing comments on a draft version of this paper. This work is partly supported by ERC grants 682608 and 787185 under the European Union’s Horizon 2020 programme, and partly supported by NSF grant PHY-1719924.

\bibliographystyle{ytphys}
\let\bbb\bibitem\def\bibitem{\itemsep4pt\bbb}
\bibliography{ref}

\end{document}